\documentclass[universe,review,submit,moreauthors,10pt,a4paper]{mdpi} 
\usepackage{amssymb}
\preto{\abstractkeywords}{\nolinenumbers}
\firstpage{1} 
\articlenumber{x}
\doinum{10.3390/------}
\pubvolume{xx}
\pubyear{2017}
\copyrightyear{2017}
\externaleditor{Academic Editor: name}
\history{Received: date; Accepted: date; Published: date}

\Title{Scalar-tensor black holes {embedded} in an expanding universe}

\Author{D.A.Tretyakova $^{1}$\orcidA{}, B.N.Latosh $^{2}$\orcidB{}}

\AuthorNames{D.A.Tretyakova, B.N.Latosh}

\address{%
$^{1}$ \quad Department Theoretical Physics, Ural Federal University, Ekaterinburg, Russia; \\
$^{2}$ \quad Department of Physics \& Astronomy, University of Sussex, Brighton, United Kingdom;}

\corres{Correspondence: daria.tretiakova@urfu.ru}

\abstract{In this review we focus our attention on scalar-tensor gravity models and their empirical verification in terms of black hole and wormhole physics. We focus on black holes, embedded in an expanding universe, describing both cosmological and astrophysical scales. We show that  in scalar-tensor gravity it is quite common that the local geometry is isolated from the cosmological expansion, so that  it does not backreact on the black hole metric. We try to extract common features of scalar-tensor black holes in an expanding universe and point out {the issues that are not fully investigated}.}


\begin{document}
%%%%%%%%%%%%%%%%%%%%%%%%%%%%%%%%%%%%%%%%%%
%% Only for the journal Gels: Please place the Experimental Section after the Conclusions

%%%%%%%%%%%%%%%%%%%%%%%%%%%%%%%%%%%%%%%%%%
%\setcounter{section}{-1} %% Remove this when starting to work on the template.
%\section{How to Use this Template}
%The template details the sections that can be used in a manuscript. Note that the order and names of article sections may differ from the requirements of the journal (e.g. the positioning of the Materials and Methods section). Please check the instructions for authors page of the journal to verify the correct order and names. For any questions, please contact the editorial office of the journal or support@mdpi.com. For LaTeX related questions please contact Janine Daum at latex-support@mdpi.com.

\section{Introduction}\label{Introduction}

General Relativity (GR) is nowadays the most successful theory of gravity providing the best fit for all the observed gravitational phenomena. For a century GR was only tested in the weak field regime \cite{Will:2014kxa,Hinshaw:2012aka,Ade:2015xua} until the recent detection of gravitational waves \cite{Abbott:2016blz,Abbott:2016nmj,Abbott:2017vtc} which provided a completely new way to study strong-field gravity \cite{TheLIGOScientific:2016src}\nocite{Moffat:2016gkd,Cayuso:2017iqc}-\cite{PhysRevLett.119.141101} via optical and gravitational channels \cite{PhysRevLett.119.161101,GBM:2017lvd}. Furthermore constantly increasing resolution of satellite and terrestrial observational facilities gives one hope to resolve the SgrA* in the near future. \cite{Fish_2016,Ortiz-Leon:2016cch}. 

However on this successful background we encounter several hints on the GR incompleteness\cite{Berti:2015itd,Capozziello:2011et} which are:
\begin{itemize}[leftmargin=*,labelsep=5.8mm]
\item dark matter,
\item dark energy,
\item inflation.
\end{itemize}
These issues can be approached in terms of the classical GR action or within the extended gravity framework \cite{Bertotti:2003rm,Capozziello:2011et} and we will now describe both ways in brief. 

Dark matter is generally treated either as new gravitational effects lying beyond GR reach or as evidences of beyond standard model particle physic. Empirical evidences favour the former case \cite{Clowe:2006eq,Boran:2017rdn}, but both approaches seem to provide an adequate description of the dark matter  \cite{Blumenthal:1984bp}-\nocite{Davis:1985rj,Capozziello:2004us,Capozziello:2006uv}\cite{Capozziello:2006ph}. 

Dark energy, on the other hand, appears to be a pure gravitational phenomenon. $\Lambda$CDM model provides the best fit for contemporary observational data \cite{Hinshaw:2012aka,Ade:2015xua}. The main problem here is that one cannot evaluate the cosmological constant from the first principles. Moreover, its value appears to be in a contradiction with the standard model: this is the cosmological constant problem \cite{Zeldovich:1968ehl,Weinberg:1988cp}.

Inflation is one of the key motivations to consider extended gravity, as it appears to be impossible to construct a self-consistent cosmological model without the inflationary phase \cite{Linde:2007fr,Senatore:2016aui}. Inflation does not occur in GR, so we are forced to consider extended gravity which offers two basic ways to account for inflation. First one is to consider quantum corrections to a matter Lagrangian in the realm of curved spacetime. Such corrections provide model renormalizability and result in pure gravitational terms. The approach is known as $f(R)$ gravity and it was first applied in the paper \cite{Starobinsky:1980te} demonstrating the appearance of an inflationary phase. Another way to inflation is to introduce a new scalar degree of freedom which is called inflaton. During the early cosmological evolution the inflaton is strongly coupled to gravity and its energy density drives the inflation \cite{Linde:1983gd}. This framework is called scalar-tensor (ST) gravity.

{While dark matter and dark energy interpretations are ambiguous, inflation provides a direct way to extended gravity: f(R) gravity or scalar-tensor one.}
By means of conformal transformations one can translate an $f(R)$ gravity model into a ST model \cite{DeFelice:2010aj,Nojiri:2017ncd,Nojiri:2010wj,Nojiri:2006ri}. We discuss this matter in Section \ref{Conformal_frame} in detail, here we would like to briefly mention that one can treat conformal transformations as a way to map {complicated form of}  $f(R)$ gravity effects with the scalar field. {Therefore ST models may  be treated as a framework to account for beyond GR gravitational phenomena by means of new gravitational terms or new scalar degrees of freedom. In our paper we therefore consider f(R) gravity mostly to be another shape of the  of the scalar-tensor theory and concentrate on the scalar-tensor formulation. We also prefer to focus of scalar-tensor models rather on $f(R)$-gravity because the former mathematical framework has a power to treat the auxiliary degree of freedom separately thereby providing a precise tool to track its influence on the gravitational phenomena. Moreover, recently discovered scalar-tensor models of the Horndeski class \cite{Charmousis:2011bf} may not be mapped on $f(R)$-gravity \cite{Momeni:2014uwa} (only on $f(R,R_{\mu\nu})$-gravity) because of the derivative coupling. Scalar-tensor gravity was also one of the firs attempts to modify general relativity \cite{Brans:1961sx}. Therefore scalar-tensor models represent the cornerstone of extended gravity and we focus on them in this review.}

{The research filed of scalar-tensor gravity is vast and cannot be covered in one paper. Therefore we focus our attention on ST models empirical verification in terms of black hole and wormhole physics. The goal of this paper is to highlight that the most promising framework to test this models is given by a comprehensive approach, when one considers  a black hole in the context of the cosmological background, i.e. embedded in an expanding universe. Speaking of such a black hole we mean a self-consistent solution of the spherically symmetric field equations in vacuum in a model, describing the expanding Universe.} Such a black hole will usually have the structure of the Schwarzschild -- de Sitter solution describing both cosmological and astrophysical scales. However, as we shall see, in scalar-tensor gravity it is quite common that the local geometry is isolated from the cosmological expansion, so that  it does not backreact on the black hole metric.  We try to extract common features of scalar-tensor black holes in an expanding universe and point out {the issues that are not fully investigated}. Fulfilling the requirements above, we are restricted to a small subset of  scalar-tensor black holes, since most of the known solutions are vacuum or have no explicit analytical form due to the field equations complexity. {For a more general picture of black hole solutions in extended gravity we recommend the following reviews \cite{delaCruzDombriz:2012xy,Nojiri:2006ri,Nojiri:2010wj}.}

This paper is structured as follows. In Section \ref{ST_models} we discuss general features of ST models and their physical content. We proceed with Section \ref{SAdS} devoted to the standard Schwarzschild--de Sitter solution. In Section \ref{No-hair_theorems} we discuss no--hair theorems for ST gravity. We clarify that the absence of scalar hair constraints the black hole (but not wormhole) phenomenology and excludes any beyond Schwarzschild--anti-de Sitter black hole. Sections \ref{Brans-Dicke_theory} - \ref{stf} are devoted to specific scalar-tensor frameworks, e.g. Brans--Dicke and Horndeski theories. We conclude in Section \ref{concl}.

%\section{{Extended gravity black holes}}
%
%{A number of complimentary reviews of black hole in modified gravity should also be appreciated \cite{delaCruzDombriz:2012xy,Nojiri:2006ri,Nojiri:2010wj}.}

\section{Scalar-tensor gravity}\label{ST_models}

ST models occupy a huge branch of extended gravity and play a central role in gravity research. These models originate from Brans--Dicke theory \cite{Brans:1961sx} initially created to incorporate Mach's principle into the GR:

\begin{equation}\label{Brans-Dicke_action}
S_\text{BD}=\cfrac{1}{16\pi} \int d^4 x ~ \sqrt{-g} \left[ \phi \mathcal{R} - \cfrac{\omega}{ \phi} \partial_\mu \phi \partial^\mu \phi \right],
\end{equation}

where $\phi$ is the Brans--Dicke scalar field, which manifests itself thorough the variation of the gravitational constant, $R$ is the scalar curvature, and $\omega$ is a constant. The following generalization of Brans--Dicke action \eqref{Brans-Dicke_action} is often considered as a general form of a ST model Lagrangian \cite{Bergmann:1968ve,Wagoner:1970vr,Nordtvedt:1970uv}:

\begin{eqnarray}\label{general_scalar-tensor_action}
 S_\text{ST}=\cfrac{1}{16\pi} \int d^4 x ~\sqrt{-g} \left[ \phi R - \cfrac{\omega(\phi)}{\phi} \partial_\mu \phi \partial^\mu \phi - 2 \Lambda(\phi) \right].
\end{eqnarray}

Here $\omega(\phi)$ is the coupling parameter and $\Lambda(\phi)$ is a $\phi$-dependent generalization of the cosmological constant, i.e the scalar field potential. Multiple ST gravity theories were developed in the past years and new ones emerge constantly (see Section 3.1 of \cite{Clifton:2011jh} and Section 2.2 of \cite{Berti:2015itd} for detailed reviews of ST gravity). Through the years they were all subject to theoretical and observational tests and Brans--Dicke theory among them was tested with the greatest care. The sole result is the bound on the model parameter $|\omega|>50000$ imposed via the parametrized post-Newtonian (PPN) expansion of the theory \cite{Bertotti:2003rm,Hohmann:2013rba} (even $|\omega|>10^{10}$ according to \cite{Deng:2016moh}, and the paper \cite{Tretyakova2011ch} proposes $|\omega|>10^{40}$).

The scalar field and related values enter field equations and affect the space-time geometry. In what follows we briefly point out the main test beds for scalar--tensor gravity.

At the scale of stellar systems the scalar field changes both internal structure and external gravitational field of a star. An example is the spontaneous scalarization process \cite{Damour:1993hw,Damour:1996ke} occurring in massive neutron stars in the strong field regime. The scalar field receives an effective position-dependent mass due to its nonlinear coupling to matter thereby affecting the star internal structure. Spontaneous scalarization was demonstrated both in analytical \cite{Damour:1993hw,Damour:1996ke} and numerical studies \cite{Shibata:1994qd,Harada:1996wt,Novak:1997hw}. This allows one to confront ST theories with observations by modelling the stellar interior. 

At the level of black holes the scalar field can violate no--hair argument and produce scalar hairs. We discuss this issue in Section \ref{No-hair_theorems}, here we would like to mention that a scalar field is an essential component for a no--hair theorem violation, as it can bifurcate the theorem requirements in a healthy manner by means of the nonlinear coupling with gravity. The wormhole is also a quite common result of such bifurcation. The scalar field plays a role of an exotic matter supporting a wormhole throat thereby opening a new class of gravitational phenomena. {Moreover, new scalar degree of freedom  affects the black hole thermodynamics. One cannot simply adopt Bekenstein-Hawking formula $S=1/4 A$, as scalar field may bring an additional contribution on black hole entropy. We discuss this issue in more detail in Sections \ref{Brans-Dicke_theory} and \ref{BD_with_potential}.}

The scalar field also manifests itself at the scales of galaxies and galaxy clusters. In an expanding universe the size of a gravitationally coupled object is defined by the balance of the gravitational attraction and the force expanding the universe. The maximal size of a gravitationally coupled object is known as a turnaround radius (see also zero-velocity surface) \cite{Faraoni:2015saa,Faraoni:2015zqa}. The scalar field affects the value of the turnaround radius, as it contributes to the total amount of energy density in the area \cite{Pavlidou:2013zha,Pavlidou:2014aia,Tanoglidis:2014lea}. The value of the turnaround radius can be established using the galaxy dust density profile which experiences the leap at the border of a gravitationally coupled area. Contemporary methods do not yet allow one to measure this leap rigorously, but some constraints on model parameters can nevertheless be placed \cite{Pavlidou:2013zha}.

Finally, ST models trigger new phenomenology at the cosmological scale \cite{Tretyakova:2011ch}\nocite{Damour:1992kf,Elizalde:2004mq}-\cite{Ratra:1987rm}. Contributing to the matter stress-energy tensor, the scalar field affects the cosmological dynamics and ST theories can thereby be tested using the entire cosmological dataset \cite{Demianski:2007mz, doi:10.1142/S2010194511000250, PhysRevD.95.063502}.

Some of ST models are equipped with the Vainshtein mechanism \cite{Babichev:2013usa}, which screens out the scalar field due to its nonlinear coupling to matter. In the high density area the scalar field particles interact strongly with matter and do not propagate prohibiting scalar field influence. As a result of screening around localized objects, a transition region must appear where the solution interpolates between the cosmological and the local values of the parameters. This means that the local solution does not feel the cosmological scale properly and at the same time the long-range cosmological solution should not be read-off directly from the large $r$ hevahior of the metric, as it can be done in the GR \cite{Bellini:2014fua}. However, Vainshtein mechanism requires a specific form of the scalar field interaction to take place and cannot wipe out the scalar field influence in the low matter density regime. Therefore one may expect beyond GR effects even in models with the screening mechanism.

Most of the research papers attempting to test ST gravity models consider a single selected scale (and hence a single observational dataset). An extended gravitational theory must provide a uniform description for the phenomena taking place at all scales. We would like to focus our attention on the role of a comprehensive approach to both small (though astronomical) and large scale physics in ST gravity. To do so we consider models, where such an approach can result in valuable constraints and compare bounds imposed by cosmological, astrophysical and weak field observations. A commonly emerging class of solutions describing an object in an expanding universe is the asymptotically anti-de Sitter solution (AdS) so we concentrate on such solutions hereafter. 

\subsection{Conformal frame}\label{Conformal_frame}

Before proceeding to spherically symmetric solutions we would like to make some comments on choosing the conformal frame. 

One can use conformal transformations to map an $f(R)$ gravity model onto a ST one in the same way one can map one ST model onto another. Such mapping does not preserve a form of matter gravitational interaction. The frame in which matter only interacts with metric is called the Jordan frame, but in that frame the interaction between the scalar field and gravity may acquire a non-trivial form. The frame possessing the standard Hilbert-Einstein Lagrangian is called the Einstein frame, but in that frame the scalar field may interact with matter perturbing its propagation along the geodesics. Despite these frames are mathematically equivalent at the level of the classical action, the discussion on their physical equivalence is still going. Some authors claim that Jordan frame is not physically equivalent to Einstein one because of the energy conditions violation \cite{Magnano:1993bd,Faraoni:1999hp}. Others state that these frames are equivalent because one should compliment a conformal transformation with a unit transformation thereby neglecting the apparent difference \cite{Postma:2014vaa,Catena:2006bd,Faraoni:2006fx}. Cauchy problem is well-stated both in Jordan and Einstein frames \cite{Salgado:2005hx}.

% The comment to Referee No 2
{In the context of conformal equivalence one also should consider conformal transformations of observable quantities, as a physical quantity available for measurement may not be invariant under the transformations. For instance, spacetime singularity may have a different form in Jordan and Einstein frames \cite{Bahamonde:2016wmz}. Another example is provided by  \cite{Bahamonde:2017kbs} where it is proven that accelerated expansion of the universe in one model may be perceived as a decelerated expansion in the other. Therefore one must choose a consistent method to relate observable quantities with the experimental data in the context  of conformal frames equivalence. }
Paper \cite{Magnano:1993bd} presents the brief overview of the issue.

The situation is similar in the realm of quantum theory. In paper \cite{Calmet:2012eq} the authors demonstrated that Jordan and Einstein frames are equivalent within the path integral approach, and the paper \cite{Pandey:2016unk} asserts that Jordan and Einstein frames of Brans--Dicke theory are equivalent at the quantum level. At the same time the first loop contribution differs in some ST models for Jordan and Einstein frames \cite{Kamenshchik:2014waa}.

In this review we only consider classical implications of ST models, so we omit the issue of frame equivalence at the quantum level. At the classical level we treat Jordan and Einstein frames as complimentary frameworks to map complex gravitational effects.

\section{Schwarzschild--anti-de Sitter solution}\label{SAdS}

We now recall general features of the black hole in the de Sitter universe. 
According to the Birkhoff theorem spherically symmetric solutions of Einstein vacuum field equations with $\Lambda \neq 0$ reduce to Schwarzschild-de Sitter (SAdS) metric \cite{kottler}:

\begin{eqnarray}
ds^2 &=& U(r) ~ dt^2 - \cfrac{dr^2}{U(r)} -r^2 d\Omega^2 ~,\\
U(r) &=& 1 - \cfrac{2 m }{r} -\cfrac{\Lambda}{3} ~ r^2 ~,
\end{eqnarray}

where $\Lambda>0$ is the cosmological constant driving the large-scale expansion, and $m$ is the black hole mass. Definitions of mass and angular momentum are complicated because of the asymptotic structure of de Sitter spacetime, the detailed discussion is given in \cite{Ashtekar:1984zz}. Solution stability w.r.t linear perturbations was proven in \cite{Kodama:2003jz}, its thermodynamic properties were studied in \cite{Hawking:1982dh}. The turnaround radius is given by the following expression:

\begin{eqnarray}
r_\text{t} = \sqrt[3]{\cfrac{3m}{\Lambda}} ~.
\end{eqnarray}

The equation above relates the black hole mass and the cosmological constant thereby showing connection between small and large scale physics and providing a uniform description for diverse physical phenomena. A similar feature occurs in thermodynamics: the black hole cannot have a temperature below the critical one

\begin{eqnarray}
T_C = \cfrac{1}{2\pi} \sqrt{\Lambda},
\end{eqnarray}

because it is embedded into a thermal radiation background generated by the cosmological horizon. At the same time the thermal radiation with the temperature below $T_C$ cannot form a black hole. {Despite the existence of a minimal temperature, Bekenstein-Hawking formula for entropy holds. The laws of black hole thermodynamics in a de Sitter space resemble the standard features of a black hole in an asymptotically flat space \cite{Hawking:1982dh}.}

\section{No--hair argument}\label{No-hair_theorems}

The no--hair argument plays a central role in black hole physics, constraining the number of non-trivial configurations available (see \cite{Sotiriou:2015pka} for a recent review of no--hair theorems). A model admitting the no--hair theorem is indistinguishable from GR at the phenomenological level. Therefore a violation of the no--hair argument is required to draw some conclusions on the extended gravity model by means of black hole observations. Here we would like to discuss cornerstones of the argument.

The Hawking theorem \cite{Hawking:1971vc} states that a rotating stationary black hole is axisymmetric and has a topologically spherical boundary. For a ST model this means that the scalar field is constant and the resulting solution is indistinguishable from the standard GR one. In ST models with a Lagrangian \eqref{general_scalar-tensor_action} one can perform a conformal transformation to the Einstein frame. The resulting vacuum field equations are identical to the Einstein equations in the presence of the scalar field and this is where the Hawking theorem applies.  Three main assumptions of the no--hair argument are crucial in our context. 

First, one assumes that the spacetime is asymptotically flat or de Sitter. In Brans--Dicke theory the cosmological expansion cannot be realised without the cosmological constant or the scalar field potential. Therefore Brans--Dicke model with negative cosmological constant is constrained by the no--hair argument and provide no new phenomenology at the black hole level. Note that despite no--hair argument holds for an asymptotically de Sitter spacetime, it should be violated in the spacetime with anti--de Sitter asymptotic \cite{PhysRevD.64.044007}. 

One should also exclude nontrivial interactions between the scalar field and gravity. Scalar field interacting with Gauss-Bonnet term, which we discuss in Section \ref{ST_with_GB-term}, provides an example of such interaction \cite{Sotiriou:2013qea,Sotiriou:2014pfa}. Gauss-Bonnet contains a coupling between the Kretschmann scalar $R_{\mu\nu\alpha\beta} R^{\mu\nu\alpha\beta}$, vanishing only in the flat spacetime, and the scalar field. Gravity in the strong field regime induces the strong field regime on the scalar field and forms scalar hairs.

Finally, hair may develop due to the energy conditions violation. This can be achieved in the phantom as well as in the regular regime of the scalar field \cite{Elizalde:2004mq}. Therefore more general theories, like \eqref{general_scalar-tensor_action} admit hairy asymptotically de Sitter black holes or wormholes, induced by the energy conditions violation.

The no-hair argument was layed out for the most of the theories in question, however there is usually a way to violate it's without imposing any extraordinary requirements.

\section{Brans--Dicke theory}\label{Brans-Dicke_theory}

Spherically symmetric solutions within the Brans--Dicke theory \eqref{Brans-Dicke_action} were obtained in \cite{Brans:1961sx,Brans:1962zz} and divided into four classes connected by nontrivial transformations \cite{Bhadra:2001fx}--\nocite{Bhadra:2005mc,Wyman:1981bd}\cite{Buchdahl:1959nk}. For a long time they were thought to be black holes, however dependent on the value of $\omega$ they describe either a wormhole (for the phantom scalar) or a naked singularity (for a normal non-trivial scalar field) \cite{Agnese:1995kd,Faraoni:2016ozb}. Regular black hole geometry can only be achieved in the GR limit, when the scalar field $\phi$ is constant and $\omega =\infty$ (see \cite{Bhattacharya:2015iha} for a detailed study). Thus if we are interested in the nontrivial geometry, yielding measurable effects and allowing us to test Brans-Dicke theory, we are forced to deal with wormholes.

{Thermodynamic properties of the original Brans-Dicke black hole solutions \cite{Brans:1961sx,Brans:1962zz} were studied in \cite{Kim:1996fm}. One cannot apply Bekenstein-Hawking entropy formula directly, as the scalar field can bring an additional entropy to the black hole (see \cite{Faraoni:2010yi}and references therein). The black hole entropy is related with the temperature as follows:}

\begin{eqnarray}
  S = [2-(Q+3 \chi)] ~ \cfrac{r_0}{T_H},
\end{eqnarray}

{where $T_H$ is the Hawking temperature, $r_0$ is the horizon radius, and $Q$ and $\chi$ are solution parameters. A detailed study of the black hole thermodynamics \cite{Kim:1996fm} revealed that only solutions with $Q=1$, $\chi=0$ are physically relevant, as other configurations have zero entropy and infinite temperature. }

{On the more general ground one can proof that the black hole entropy in Brans-Dicke theory is given by the following \cite{Kang:1996rj}:}

\begin{eqnarray}
  S = \cfrac14 \int\limits_\Sigma d^2 x \sqrt{g^{(2)}} \phi , \label{BD_entropy}
\end{eqnarray}

{where $\Sigma$ denotes integration over black hole horizon, $\phi$ is the scalar field and $g^{(2)}$ is the metric on the horizon. In \cite{Kang:1996rj} it was also proven, that the entropy remains constant under the conformal transformation. The fact that integral \eqref{BD_entropy} should be evaluated at the horizon puts additional constraints on the solution. If the scalar field diverges at the horizon, one cannot attribute it a finite entropy, so the solution should be considered as unphysical. Solutions with the vanishing scalar field at the horizon are known as cold black holes and they are subjected to a careful scrutiny \cite{Zaslavskii:2002zv,Nandi:2000gt,Bronnikov:1998hm}.}

Wormholes were initially designed as an object \cite{Morris:1988cz} embedded in a static universe and supported by exotic matter\footnote{We call exotic matter the one that violates the null energy condition (NEC)}. Nonetheless one can create a wormhole solution embedded in a dynamical universe in the presence of a phantom matter\footnote{Phantom matter violates the strong energy condition} \cite{Morris:1988cz, Heydarzade:2014ada,Lemos:2003jb}. The exterior gravitational field of a wormhole matches the (anti-)de Sitter spacetime far from the throat. A common feature of such wormholes is the time evolution of the throat, namely, a traversable wormhole may become un-traversable through the universe evolution \cite{Narahara:1994np}. Within GR one can also construct a time-dependent wormhole embedded in an anisotropic traceless fluid which does not require exotic matter until some critical time when the energy condition is violated inside the wormhole \cite{Riazi:2000uu}. 

Before we proceed with describing wormholes in Brans-Dicke theory we must give a few comments on the energy conditions violation normally needed to support a wormhole. Within Brans-Dicke theory such a violation is usually achieved with $\omega<0$. Negative values of $\omega$ are often ruled out because of the wrong sign before the kinetic term in the action leading to a ghost in the quantum theory (and to a phantom scalar field in cosmology \cite{Elizalde:2004mq}). Nevertheless researchers do consider $\omega<0$ for the following reasons. Modern observational data indicate that the phantom nature of dark energy is more likely \cite{2013ApJS19H}. Combining WMAP+eCMB+BAO+H0+SNe data yields the equation of state (EOS) parameter\footnote{The relation of energy and pressure density.} of the dark energy $w_{DE_0} = -1.17(+0.13 -0.12)$ for the flat Universe at the significance level of $68\%$. This range of $w_{DE_0}$ favours the phantom nature of the dark energy. Similar picture occurs for the non-flat universe model and for different dataset combinations, e.g. $w = -1.019 ~\begin{matrix} +0.075 \\ -0.080 \end{matrix}$ according to the PLANK data \cite{Will:2014kxa,Hinshaw:2012aka,Ade:2015xua}. In the classical Brans-Dicke theory phantom regime can be achieved with $\omega<0$. Finally, consideration of a ghost scalar field as an effective description for a theory with positive defined energy could eliminate quantum contradictions \cite{Nojiri:2003vn, Carroll2003st}. All these arguments make the consideration of Brans-Dicke theory with $\omega<0$ reasonable. Modern observational bounds only limit the absolute value of $\omega$ and modelling the gravitational collapse does not rule out negative $\omega$ values as well \cite{PhysRevD.51.4236}. So the range $\omega<0$ is of great interest from the phenomenological point of view, since it can provide no-ghost but phantom cosmology in agreement with modern observations (see \cite{Elizalde:2004mq} for a no-ghost but phantom cosmological model). Therefore we proceed with considering wormhole solutions on a cosmological background of the Brans--Dicke theory.

In the classical Brans--Dicke formalism one can construct static as well as dynamic wormholes. The paper \cite{Riazi:2000uu} (see also references therein) introduced a time-dependent Brans-Dicke wormhole solution in an evolving cosmological background without the cosmological constant. Geometry is supported by a traceless matter, which in some cases can be non-exotic throughout space. The background spacetime conforms with the well-known radiation-dominated Friedmann model and does not describe the de Sitter expansion. For the solution to be non-trivial one has to violate the weak energy condition by means of negative values of $\omega$, the resulting wormhole evolves along with the universe. This is a common feature for wormhole solutions embedded in a cosmological background. 
To describe the wormhole embedded in an expanding universe the solution should possess de Sitter asymptotic, so we turn to such solutions now. 

\subsection{Brans--Dicke theory with the non-vanishing cosmological constant}\label{BDL}

Brans-Dicke theory successfully describes an expanding universe by means of the scalar field potential or the cosmological constant (BD$\Lambda$), i.e. $\omega = \text{const}$, $\Lambda (\phi) = \text{const} \times \phi$ \cite{Uehara:1981nq}--\nocite{Riazi:1995ms,Pimentel1985,Ram:1997un,Pandey:2001ua,Romero:1992ci,Romero:1992xx,Cervero1983}\cite{Tretyakova:2011ch}. However for the best of our knowledge, {\bf no wormhole solutions within this framework were presented in the literature}.

The value of a turnaround radius for BD$\Lambda$ was calculated in \cite{Bhattacharya:2016vur}. Authors considered two different approaches: they studied spherically symmetric gravitational field of a point mass and cosmological perturbations. Calculations resulted in the following analytical expression for the turnaround radius:

\begin{eqnarray}
 r_\text{t} = \sqrt[3]{\cfrac{3 m}{\Lambda} \cfrac{2\omega+4}{2\omega+1}} = \left. r_\text{t} \right|_\text{GR} \times \left( 1+ \cfrac{1}{2\omega} + O\left(\cfrac{1}{\omega^2}\right)\right) ~.
\end{eqnarray}

The standard GR expression for the turnaround radius is recovered in $\omega=\infty$ limit, receiving positive corrections suppressed by the $\omega^{-1}$ factor. According to \cite{Bhattacharya:2016vur} this correction is consistent with the contemporary observational data.

\subsection{Brans-Dicke theory with the scalar field potential}\label{BD_with_potential}

{\bf Cosmology via the scalar field potential}

A variety of cosmological phenomena in ST models with an arbitrary scalar field potential is much wider that in BD$\Lambda$. Moreover, such models are not excluded by the observations and serve as the cornerstone of inflation.

The simplest example is given by quintessence models. One can treat the scalar field as matter alongside with the standard model particles. In that case the EOS parameter $w$ is given by the following:

\begin{eqnarray}
 w = \cfrac{\rho}{p} = \cfrac{\dot{\phi}^2 - 2 V }{ \dot{\phi}^2 + 2V},
\end{eqnarray}

where $\dot \phi$ is a scalar field derivative w.r.t. time and $V$ is the scalar field potential. In such a setting $w$ is constrained to lie between $-1$ and $1$, which is allowed by the observational data. The quintessence scenario appears in models with a negatively defined scalar field potential which satisfies the contemporary cosmological dataset.

{\bf Wormhole solutions}

A wormhole solution for the Brans-Dicke setting with the $\Lambda \phi^2$ potential is given in \cite{Xiao:1991nv}. It is obtained for $\phi=\phi(t)$ and the resulting wormhole is non-static, its throat radius increases as the cosmic time increases. The wormhole connects two asymptotically flat regions and does not describe an expanding universe in the background. The situation is interesting, since the de Sitter expansion occurs in the corresponding cosmological model (and even a bounce for $\omega<-3/2$) \cite{Hrycyna:2013yia}, but does not show itself in the wormhole solution.
However, \cite{Xiao:1991nv} gives no explicit solution, therefore the ability of the metric to describe the real world is unclear.

Another non-trivial model with an exponential potential was considered in \cite{Elizalde:2004mq}

\begin{eqnarray}\label{Elizalde_model}
S = \cfrac{1}{16\pi} \int d^4 x \sqrt{-g} ~e^{\alpha\phi} \left[ R - \omega \partial_\mu \phi \partial^\mu \phi -V_0 \exp\left[ \phi/\phi_0\right] \right],
\end{eqnarray}

where $\alpha$, $\omega$ and $\phi_0$ are constants. Performing the transformation to the Einstein frame one can see that even if $\omega$ is negative, in the case when 

\begin{equation}
3\alpha^2/2+\omega >0 \label{noghost}
\end{equation}

the effective kinetic energy of $\phi$ remains positive and the ghost does not appear in spite of the phantom cosmology in the background \cite{Elizalde:2004mq}. This is a highly desirable feature, relaxing the main objection against the cosmological phantom scalar field, i.e. the ghost appearance. The authors also claim that quantum effects may prevent the finite-time future singularity associated with the phantom.

The paper \cite{Tretyakova:2015vaa} considers spherically symmetric solutions in the framework above. For the following combination of model parameters:

\begin{eqnarray}\label{Constrait_1}
\phi_0 + \cfrac{\alpha}{2 (\alpha^2 + \omega)} =0.
\end{eqnarray}

the metric obtained by Agnese and La Camera \cite{Agnese:2000dv} solves the field equations:

\begin{eqnarray}
ds^2 = A^m(r) dt^2 - A^{-n}(r) dr^2 -A^{1-n}(r) ~r^2 (d\theta^2 + \sin^2\theta d \varphi^2) ,
\end{eqnarray}

\begin{eqnarray}
A(r) = 1-\cfrac{2\eta}{r} ,& m = \sqrt{2/(1+\gamma)} , & n = \gamma\sqrt{2/(1+\gamma)} .
\end{eqnarray}

Parameter $\eta$ eta relates to the solution mass $M$ as:

\begin{eqnarray}
\eta = M \sqrt{(1+\gamma)/2}.
\end{eqnarray}

The PPN parameter $\gamma$ enters the metric and provides a direct way to relate it with observational data. Expressions for $\alpha$, $\omega$ and $\phi$ are given in \cite{Tretyakova:2015vaa}. If we suppose that the scalar field potential acts as the cosmological constant, we can arrive at the conclusion that the accelerated expansion of the universe does not contribute to local dynamics, since the metric is exactly the same as in the classical Brans-Dicke setting. This effect is analogous to \cite{Xiao:1991nv} and indifferent to the value of the cosmological constant. This property agrees with the fact that the cosmological expansion does not manifest itself in the Solar system. Astrophysical properties were studied by Alexeyev et al. in \cite{Alexeyev:2011hc} and were found to be in agreement with modern observations. Therefore the metric above represents a valid candidate for a real-world object.

Following the scope of our paper we complement the above with the weak field limit study. Evaluating the PPN parameter $\gamma$ in terms of model parameters and applying the corresponding observational bound we obtain:

\begin{eqnarray}
\alpha^2 /\lvert\omega\rvert < 10^{-5},
\end{eqnarray}

where $\omega$ should be negative. Despite the model still provides a valuable description of compact objects, the constraint above comes in a contradiction with the no-ghost condition, thereby nullifying the main advantage of the model \cite{Tretyakova:2015vaa}. Both cosmological and astronomical solutions assert versatile bounds on model parameters, that resulted jointly in a more rigorous test of the model.

To sum up, the non-trivial phenomenology on the Brans--Dicke cosmological background requires unnatural scalar field configurations, however the question regarding BD$\Lambda$ is still open. 

\section{Horndeski theories}
The most general scalar-tensor action resulting in second order field equations (i.e. avoiding a ghost related to higher order terms) was proposed by Horndeski \cite{Horndeski:1974wa}. 
The Brans--Dicke theory, considered before, appears to be a narrow subclass of Horndeski models. In its modern reformulation, Horndeski theory is written as a generalized galileon Lagrangian,

\begin{eqnarray} \label{ssgg}
&&L = L_2+L_3+L_4+L_5 ,
\\
&&L_2 = G_2 , \nonumber
\\ 
&&L_3 = -G_3 \Box \phi , \nonumber
\\
&&L_4 = G_4 R + G_{4X} \left[ (\Box \phi)^2 -(\nabla_\mu\nabla_\nu\phi)^2 \right] , \nonumber
\\
&&L_5 = G_5 G_{\mu\nu}\nabla^\mu \nabla^\nu \phi - \frac{1}{6} G_{5X} \big[ (\Box \phi)^3 - 3\Box \phi(\nabla_\mu\nabla_\nu\phi)^2 + 2(\nabla_\mu\nabla_\nu\phi)^3 \big] , \nonumber 
\end{eqnarray}

where $G_2$, $G_3$, $G_4$, $G_5$ are arbitrary functions of $\phi$ and $X=- \partial^\mu \phi \partial_\mu \phi/2$, the canonical kinetic term, $R$ is the Ricci scalar, $G_{\mu\nu}$ is the Einstein tensor, and

\begin{eqnarray}
\\
&& (\nabla_\mu\nabla_\nu\phi)^2 = \nabla_\mu\nabla_\nu\phi \nabla^\nu\nabla^\mu\phi, \nonumber 
\\
&& (\nabla_\mu\nabla_\nu\phi)^3=\nabla_\mu\nabla_\nu\phi \nabla^\nu\nabla^\rho\phi \nabla_\rho\nabla^\mu\phi.
\end{eqnarray}

The scalar field has the property of admitting a special symmetry in flat (nondynamical) spacetime for $G_2\sim G_3 \sim X$ and $G_4\sim G_5 \sim X^2$ , which resembles the Galilean symmetry,therefore the name galileon ~\cite{Nicolis:2008in}. 
Galileon symmetry is broken for a curved background and for general choice of $G_i$, the corresponding scalar is then refereed to as the ``generalized'' galileon~\cite{Deffayet:2009wt}. 
It is important to note that generalized galileon theory and Horndeski theory do not start from the same principle, but turn out to be identical \cite{Kobayashi:2011nu}. 

We will consider the following ansatz:

\begin{equation}
ds^2 = h(r)dt^2 - \frac{dr^2}{f(r)} - r^2d \Omega^2.
\label{ds^2}
\end{equation}

The class of Horndeski models is wast and cannot be explored for the spherically symmetric solutions in it's general form. Such solutions were found for particular subclasses of Horndeski and we review them now. 

\subsection{Shift-symmetric subclass}\label{Shift-symmetric_Horndeski}
When equipped with a remnant of the Galileon symmetry in the flat space-time \cite{Deffayet:2009mn, Deffayet:2009wt}, namely a shift symmetry for the scalar field $\phi \to \phi +c$ (with c arbitrary real constant) in arbitrary curved spacetime, the action for Horndeski/Galileons scalar-tensor model reads

\begin{equation}
S=\int d^4x\sqrt{-g}(\mathcal{L}_2+\mathcal{L}_4), \label{lag}
\end{equation}

where

\begin{eqnarray}
\mathcal{L}_2&=& G_2(X), \\
\mathcal{L}_4&=& G_4(X)R +G_{4X}[(\Box\phi)^2-(\nabla_{\mu}\nabla_{\nu}\phi)^2].
\end{eqnarray}

Here $R$ is the scalar curvature, $X=-(\nabla_\mu \phi)^2/2$ is the canonical kinetic term of the scalar field, $G_2$ and $G_4$ are arbitrary functions of $X$, and $G_{4X}\equiv d G_4/d X$.
This theory is invariant under the shift~$\phi \to \phi + c $ (with $c$ being an arbitrary real constant) and the reflection~$\phi\to-\phi$.

For the choice of the arbitrary functions

\begin{equation}
G_2(X)=-2\Lambda+2\eta X,~~~G_4(X)=\zeta+\beta X,~~~G_3(X)=G_5(X)=0, \label{dc2Gmn}
\end{equation}

the action can be expressed in the form~\cite{Kobayashi:2014eva}

\begin{equation}
S = \int dx^4 \sqrt{-g} \left( \zeta R - \eta \left( \partial \phi \right)^2 + \beta G^{\mu\nu} \partial_\mu \phi \partial_\nu \phi - 2\Lambda \right) \label{ac},
\end{equation}

here $ G^{\mu\nu} $ is the Einstein tensor, $ \phi $ is the scalar field, $ \zeta>0, \eta$ and $\beta$ are model parameters. Though Horndeski model is not exhausted by this action we will restrict our consideration to it, since most of the static spherically symmetric solutions we obtained within this framework. 

The model \eqref{ac} is known to admit a rich spectrum of cosmological solutions (see \cite{Starobinsky:2016kua} and references therein) describing the late-time acceleration and an inflationary phase. Moreover, for $\eta\neq 0$ it admits solutions for which the $\Lambda$-term is totally screened, while the metric is not flat but rather de Sitter with the Hubble rate proportional to $\eta/\beta$. It offers an exciting opportunity to describe the late time cosmic acceleration while screening the vacuum $\Lambda$-term and hence circumventing the cosmological constant problem. Henceforth the model is very interesting on the cosmological level.

Field equations of \eqref{ac} can be integrated completely in static spherically symmetric sector \cite{Charmousis:2015aya} once the scalar field linear time dependence $\phi=\psi(r) +qt$ is introduced. The scalar field linear time evolution seems like a natural feature on a cosmological background and can be considered as an approximation for a more general theory. Although the Vainshtein screening mechanism is generally at work in Horndeski gravity, in the case of a minimal coupling of the scalar field to matter no screening radius can be posed, so the solution must posses de Sitter asymptotic in Horndeski theory as well. Black hole solutions of the form \eqref{ds^2}
 were found in the series of papers \cite{Rinaldi:2012vy}-\nocite{Babichev:2013cya, Charmousis:2015aya, Babichev:2015rva, Anabalon:2013oea}\cite{Minamitsuji:2013ura} (see \cite{Babichev:2016rlq} for a comprehensive review). These solutions are governed by the master equations: 

\begin{eqnarray}
&& f(r) = \dfrac{( \beta + \eta r^2) h(r)}{\beta \left( rh(r)\right)'},\label{f0} \\
&& h(r) = -\frac{\mu}{r} + \frac{1}{r} \int\dfrac{k(r)}{\left( \beta + \eta r^2\right)}dr,\\
&&\phi(r) = qt+\psi(r), \label{phi}
\end{eqnarray}

where $ \mu$ plays a role of a mass term and $k$ should be derived by means of the following constraint equation:

\begin{eqnarray}
&& q^2\beta\left( \beta + \eta r^2\right)^2 -
\left( 2\zeta\beta + \left( 2\zeta\eta - \lambda\right) r^2\right)k + C_0 k^{\frac{3}{2}} =0. \label{k}
\end{eqnarray}

Here $C_0$ is an integration constant. By introducing a mild linear dependence in the time coordinate for the scalar field one evades the scalar field being singular for its derivative on the horizon \cite{Charmousis:2014zaa} and makes field equations  bifurcate the no--hair theorem at the same time. The shift symmetry is keeping field equations time-independent and consistent with the static ansatz. The screening of the cosmological expansion at the black hole level appears here as well, given by the stealth Schwarzschild solution 

\begin{eqnarray} 
&&f(r)  =  h(r)=1-\cfrac{\mu}{r} , \label{f1}\\
&&\Lambda=\eta=0.
\end{eqnarray}

 For this solution the spatial part of the scalar is frozen an it only varies in time. By using different parameter combinations the master equations  can be integrated to give various solutions, among which is the Schwarzschild--de Sitter one 

\begin{eqnarray} \label{sdss}
&& f(r)  =  h(r)=1-\cfrac{\mu}{r} +\cfrac{\eta}{3\beta}r^2 , \label{f1}\\
&& q^2=(\zeta\eta+\beta\Lambda)/(\beta\eta), C_0=(\zeta\eta-\beta\Lambda)\sqrt{\beta}/\eta.
\end{eqnarray}

with the effective cosmological constant being $\eta/3\beta$. 

Stability analysis however showed that black holes described by the action \eqref{ac} can only be stable for $q=0$ in the anti de Sitter regime, when the effective cosmological constant is negative (see \cite{Tretyakova:2017lyg} and references therein) except for solution \eqref{sdss}. With $q=0$ the stealth Schwarzschild solution above becomes trivial with the frozen scalar field and displays therefore no phenomenology beyond the GR.

{Not much is known about black hole thermodynamics in Horndeski models. As Lagrangians \eqref{ssgg} depend on the arbitrary functions, the thermodynamic analysis is complicated. A simple case $G_4=\cfrac{1}{16\pi G}$, $G_5=\operatorname{const}$ was discussed in \cite{Miao:2016aol}, although the authors simply adopted the Bekenstein-Hawking formula, without deriving it. In the paper \cite{Feng:2015wvb}, on the contrary, the authors used Wald formalism to derive relation between the black hole entropy and the horizon area. Just as in the Brans-Dicke case, the black hole entropy obtains an additional contribution from the scalar field. Altough no direct constraints can be put from  \cite{Feng:2015wvb}, it seems possible to constraint a rage of Horndeski models parameter by studying the black hole viscosity, as it is argued in \cite{Feng:2015oea}. Black hole entropy $S$ and viscosity $\eta$ are connected as  $\eta/S=1/4\pi$ a(see \cite{Feng:2015oea} and discussion there). Therefore any black hole solutions must satisfy the viscosity relation to be considered physically relevant.}

Wormhole solutions in the theory \eqref{ac} were presented in \cite{Sushkov:2017ueq, Korolev:2014hwa}.  It is shown that wormholes exist only if $\eta=-1$ (phantom case) and
$\beta > 0$.  The wormhole throat then connects two anti-de Sitter spacetimes and the solution therefore is of no observational interest, since the model does not admit the Vainshtein screening mechanism.

\subsection{Scalar-tensor models with Gauss-Bonnet term}\label{ST_with_GB-term}

The second family of solutions in the Horndeski framework admitting hairy black holes involves the Gauss-Bonnet term G

\begin{eqnarray}
\mathcal{L}_{GB}= \frac{M_P^2}{2}\left[ R - \frac12 \partial_\mu \phi \partial^\mu \phi +\alpha \phi \hat G \right], \\
\hat{G}=R_{\mu \nu \rho \sigma}R^{\mu \nu \rho \sigma} - 4 R_{\mu \nu} R^{\mu \nu} +R^2
\end{eqnarray}

The Gauss-Bonnet term is a topological invariant and thus only contributes to the equations of motion in four-dimensional spacetime when coupled to the scalar. This model is a part of the G-inflation framework \cite{Kobayashi:2011nu,Kobayashi:2010cm}. The action naturally supports de Sitter solutions equipped with an exit mechanism to a linearly expanding phase \cite{Kanti:2015pda, Fomin:2017vae}.

Sotiriou and Zhou insist that any Horndeski theory where this Gauss-Bonnet term is not forbidden must have hairy black holes \cite{Sotiriou:2014pfa, Sotiriou:2013qea}. They present a black hole for the time-independent scalar to the above theory using numerical and perturbative methods. The resulting solution contains the finite radius singularity which can be hidden behind the horizon if the black hole is massive enough. The authors also noted that corrections to observables in this model are expected to be small w.r.t. GR. However the solution was only explored in the vicinity of the horizon and it's asymptotic hevahior is unclear.

\subsection{Quartic Horndeski square root Lagrangian}

The paper \cite{Babichev:2017guv} considers black holes for the static and time-dependent scalar for the following action:

\begin{equation}
S = \displaystyle\int{\mathrm{d}^4x \sqrt{-g} \left\{ \left[\zeta + \beta \sqrt{(\partial \phi)^2/2}\right] R - \dfrac{\eta}{2} (\partial \phi)^2 - \dfrac{\beta}{\sqrt{2(\partial \phi)^2}} \left[(\Box \phi)^2-(\nabla_\mu \nabla_\nu \phi)^2\right] \right\}}.
\label{eq:action}
\end{equation}

The resulting static scalar field solution takes the following form:

\begin{equation}
f(r) = h(r) = 1 -\dfrac{\mu}{r}-\dfrac{\beta ^2}{2 \zeta  \eta  r^2}, \label{sqrbh}
\end{equation}

The solution is of the Reissner--Nordstrom form. The ``electric charge'' is not an integration constant, it only depends on the parameters of the theory. The correction would be exact same for any black hole. According to the authors for the positive $\eta$ (which would corresponding to an imaginary charge of the RN) metric, we will always have an event horizon in \eqref{sqrbh}. With a negative $\eta$ the solution behaves badly beyond the event horizon, since $\phi'$ becomes imaginary there.

The cosmological constant can be added to the initial action. The solution is then modified in the same way as it is in GR, and acquires anti-de Sitter or de Sitter asymptotic. 

\begin{equation}
f(r) =h(r)= 1-\dfrac{\mu}{r} -\dfrac{\beta ^2}{2 \zeta  \eta  r^2} -\dfrac{\Lambda}{3 \zeta} r^2,
\end{equation}

If one drops the staticity assumption,  the asymptotic flatness must be abandoned at the same time. The result would again be the Reissner--Nordstrom type solution:

\begin{equation}
h \mathop{=}_{q \rightarrow 0}1-\dfrac{\mu}{r}-\dfrac{\beta^2}{2 \eta \zeta r^2}+\dfrac{\eta q^2 r^2}{24 \zeta} + \mathcal{O}(q^4)
\end{equation}

for $\phi=qt+\psi(r)$. However the properties of the spatial part of the scalar remain and $\eta$ should be positive, and the solution is AdS in this case. The question whether this asymptotic renders the solution unphysical is open, since the we must firstly check that the theory admits no Vainshtein screening and the black hole asymptotic matters.

\subsection{Cubic Galileon}
Paper \cite{Babichev:2016fbg} studies black hole solutions in a subclass of Horndeski theory, which contains the cubic Galileon term:

\begin{equation}
S = \displaystyle\int{\mathrm{d}^Dx \sqrt{-g} \left[\zeta\: (R -2\: \Lambda)- \eta\: (\partial \phi)^2 + \gamma\: \Box \phi\: (\partial \phi)^2  \right]},
\label{eq:action}
\end{equation}

The solutions were found via perturbative methods can be interpreted as black holes immersed in a  flat, or self-accelerated universe depending on the model parameter values:

\begin{eqnarray} 
&&f(r)  =  1-\dfrac{\mu}{r}-\cfrac{\Lambda_\mathrm{eff}}{3}r^2 +\mathrm{O}(r^{-4}), \\
&&h(r)=1-\dfrac{\mu}{r}-\cfrac{\Lambda_\mathrm{eff}}{3}r^2 +\mathrm{O}(r^{-6}), 
\end{eqnarray}

The cosmological pure (A)dS solution is presented as well:

\begin{eqnarray} 
&&f(r)  =  h(r)=1-\cfrac{\Lambda_\mathrm{eff}}{3}r^2 , \\
\end{eqnarray}

The effective cosmological constant of the black hole metric is given solely by the model parameters and agrees with the one of the cosmological solution perfectly

\begin{equation}
\Lambda_\mathrm{eff} = \dfrac{\eta^2}{3 \gamma^2 } \left[\dfrac{\zeta \Lambda}{\eta} \pm \sqrt{\left(\dfrac{\zeta \Lambda}{\eta} \right)^2 - \dfrac{2 \eta \zeta}{3 \gamma^2}}\; \right]^{-1}.
\end{equation}

We can see from the above that the effective cosmological constant vanishes rendering the metric asymptotically flat for $\eta=\Lambda=0$ and dS otherwise.
The following questions remain:  does such a configuration actually take place during matter collapse and is it free of ghost, gradient and tachyon instability.

\section{Theories with auxiliary fields}

There are models which expand GR with a larger number of auxiliary fields. The spectrum of such theories is broad and includes TeVeS (Tensor-Vector-Scalar gravity), MOG(MOdified Gravity), Einstein-Aether theory, biscalar-vector-tensor gravity, and many others. In this section we focus our attention on the simplest examples of such theories namely TeVeS and MOG. Both of these theories were meant to account for the dark matter and to reproduce MOND-like mechanics in the relativistic case.

The original paper devoted to MOG \cite{Moffat:2005si} proposed a way to explain the rotation velocity curves of galaxies via new scalar and vector fields. The model is given by the following action:

\begin{eqnarray}
  S &=& \int d^4 x \sqrt{-g} \left[ \cfrac{1}{16\pi G} (R+2 \Lambda)  - \omega\left(\cfrac14 B^{\mu\nu} B_{\mu\nu} + V(\phi) \right) + \cfrac{1}{G^3} \left( \cfrac12 g^{\mu\nu} \nabla_\mu G \nabla_\nu G - V(G) \right) \right. \nonumber \\
    & & \left. +\cfrac{1}{G}\left( \cfrac12 g^{\mu\nu} \nabla_\mu\omega \nabla_\nu\omega - V(\omega)  \right) +\cfrac{1}{\mu^2 G} \left(\cfrac12 g^{\alpha\beta} \nabla_\alpha \mu \nabla_\beta \mu  - V(\mu)\right) \right] ,
\end{eqnarray}

where $\phi_\mu$ is the vector field, $B_{\mu\nu} = \nabla_\mu \phi_\nu - \nabla_\nu \phi_\mu$ is the vector field tensor, and $G$, $\omega$, and $\mu$ are three scalar fields. Auxiliary scalar and vector fields contribute attractive and repulsive Yukawa components to the Newton potential and thereby explain the dark matter phenomenology. The model appears to pass the Solar System tests \cite{Moffat:2005si,Moffat:2013sja,Iorio:2008sk,Iorio:2002jy,Iorio:2007gq}, provide  adequate cosmology  \cite{Moffat:2011rp}, and describe some galaxy phenomenology \cite{Moffat:2012wn,Moffat:2007qv} (see also \cite{Moffat:2008gi}\nocite{Moffat:2016gkd,Moffat:2006gz,Moffat:2007nj}--\cite{Moffat:2006rq}).

Black hole physics was considered \cite{Moffat:2014aja} and \cite{Mureika:2015sda}. The model contains multiple parameters which allow it to operate as a generalization of a nonlinear electrodynamics in a curved spacetime. It thereby admits black hole solutions obtained within the nonlinear electrodynamics model \cite{Moffat:2014aja,AyonBeato:1998ub,AyonBeato:1999ec}. However the authors of \cite{Moffat:2014aja} studied a solitary black hole with zero cosmological constant and flat asymptotic hevahior, therefore results of \cite{Moffat:2014aja,Mureika:2015sda} relate to the stationary universe. For the best of our knowledge, there are no papers describing the black hole in an expanding universe, therefore further research of the model is in order. The no-hair argument and the generalization of the Birkhoff theorem were also never considered, therefore the role of the solutions \cite{Moffat:2014aja,Mureika:2015sda} is not yet completely understood. {Thermodynamics of rotating and non-rotating black holes in MOG was considered in \cite{Mureika:2015sda}. General layout of MOG black hole thermodynamic is affected by the fact that some black hole solutions may posses a mass-dependent charge, the issue is discussed in great details in \cite{Mureika:2015sda}. The entropy also obtains logarithmic corrections due to quantum fluctuations.}

The situation is similar within TeVeS models (see \cite{Clifton:2011jh,Skordis:2009bf} for reviews of TeVeS in the context of modified gravity). TeVeS model was incepted in  \cite{Bekenstein:2004ne} in order to construct a relativistic MOND in terms of the nonlinear field interaction (also known as the aquadratic Lagrangian theory). The theory describes gravitational field with the physical metric $g_{\mu\nu}$ which is universally coupled to matter in order to satisfy the Einstein equivalence principle, the vector field $A_\mu$, and the scalar field $\phi$. One defines the Bekenstein metric $\widetilde{g}^{\mu\nu}$ by the following:

\begin{eqnarray}
  g_{\mu\nu}= e^{-2\phi} \widetilde{g}_{\mu\nu}-2 \sinh(2\phi) A_\mu A_\nu .
\end{eqnarray}

The metric couples to auxiliary fields and it should be used to operate with their indices. One also requires the vector field to be a time-like vector with respect to the Bekenstein metric. The model action is given by the following:

\begin{eqnarray}
  S &=& \cfrac{1}{16\pi G} \int d^4 x \sqrt{-\widetilde{g}} \widetilde{R}-\cfrac{1}{32\pi G} \int d^4 x\sqrt{-\widetilde{g}} \left[ ~K~ \widetilde{g}^{\mu\alpha} \widetilde{g}^{\nu\beta} F_{\mu\nu} F_{\alpha\beta} - 2 \lambda (\widetilde{g}^{\mu\nu} A_\mu A_\nu +1 ) \right] \nonumber \\
  & & -\cfrac{1}{16\pi G} \int d^4 \sqrt{-\widetilde{g}} \left[ \mu (\widetilde{g}^{\mu\nu} - A^\mu A^\nu) \widetilde\nabla_\mu\phi \widetilde\nabla_\nu\phi + V(\mu) \right],
\end{eqnarray}

where $\mu$ is a nondynamical dimensionless scalar field and $\lambda$ is a Lagrange multiplier.

In order to obtain the late-time cosmological acceleration one may introduce the cosmological constant as a linear term in the potential $V(\mu)$, but this only leads to the same cosmological problem, as the constant is going to be a free model parameter. At the same time, the model equipped with nonlinear field interaction which is able to produce the late-time acceleration with vanishing cosmological constant in a certain cases \cite{Zhao:2006vm}. Such phenomenology appears because the TeVeS action has a resemblance with the k-essence model \cite{ArmendarizPicon:1999rj,ArmendarizPicon:2000dh}, but we aware of no paper proving that the late-time acceleration phase take place in the model with an arbitrary set of parameters. Therefore a more detailed investigation of the model is required to establish its ability to explain dark energy phenomenology.

Despite the large number of papers devoted to black hole study \cite{Giannios:2005es,Sagi:2007hb,Skordis:2011ad}, further research is required. Black hole solution presented in \cite{Giannios:2005es} violates causality and may not be considered as a physically-relevant, in paper \cite{Sagi:2007hb} the authors constrained themself only to an asymptotically flat spacetime thereby excluding phenomenology of a black hole in an expanding universe. Only in paper \cite{Skordis:2011ad} a solution was obtained without usage of any special approximations and it appears to have a flat asymptotic. For the best of our knowledge, there are no papers giving the no-hair argument or Birkhoff theorem in TeVeS. It is also unclear if the mechanism generating the late-time acceleration in the cosmological regime \cite{Zhao:2006vm} holds for local solutions such as \cite{Sagi:2007hb,Skordis:2011ad}. { Black hole thermodynamics was studied in \cite{Sagi:2007hb} and black hole temperature and entropy appeared to be identical to the GR case.}

There are several crucial issues of models with auxiliary fields that should be clarified. First, the no-hair argument should be extended on these theories. The argument defines if a model has any significant difference from the GR at the level of black hole physics. If the no-hair argument fails to constraint this models, one should seek asymptotically de Sitter black hole solutions  to check whether the theory can provide a uniform description of small and large scale physics. Finally, one should clarify the role of an auxiliary vector field. Within the Einstein-Aether theory \cite{Blas:2014aca} the vector field leads to the violation of the Lorentz invariance, while in TeVeS gravity nonlinear coupling of scalar and vector fields can lead to the late-time acceleration phase without the cosmological constant \cite{Zhao:2006vm}. Therefore one should clarify if TeVeS violate the Lorentz invariance and how auxiliary vector field affects the scalar field phenomenology.

\section{String theory framework} \label{stf}

String theory approach to the black hole physics differs extremely from the modified gravity framework. The discussion of black holes in the string theory lies far beyond the reach of this paper (see \cite{Skenderis:1999bs,Mathur:2005zp} for a detailed review), nonetheless we present the basic features of the string theory black hole physics, since the scalar field is generally involved.

The string theory is a quantum theory describing the physical realm in terms of the  fundamental superstrings interaction. It is widely accepted that the theory provides an adequate description of the gravitational phenomena at the Planck energy scale thereby serving as a high energy gravity theory. In the low energy regime the string theory should be replaced with some effective theory for the gravitational phenomena, such as black holes, in terms of the classical spacetime. Therefore one may not be able to obtain a relevant description of a macroscopic black hole within the string theory framework, nonetheless the so-called fuzzball conjecture \cite{Susskind:1993ws,Russo:1994ev} provides a way to do this. It is well established that the classical black hole has a non-zero entropy defined by the size of its event horizon. The fuzzball conjecture states that a classical black hole with entropy $S$ should be treated as a manifestation of $\exp[S]$ black hole microstates \cite{Mathur:2005zp,Skenderis:2008qn}. This is the main reason why our method -- study of black holes with Schwarzschild-de Sitter asymptotic -- may hardly be applied in the context of the string theory. The fuzzball conjecture considers the black hole to be solitary object which is not affected by the global cosmological expansion. Therefore objects with de Sitter asymptotic lie beyond the scope of the fuzzball conjecture.

At the same time, the string theory should be replaced by an effective low energy theory for the sub-Planckian phenomena \cite{Donoghue:1994dn,Burgess:2003jk}. Such an effective theory mush describe the gravitational phenomena in terms of the classical spacetime and fields, which is exactly what modified gravity is about (see \cite{Horowitz:1996ay,Youm:1997hw,Cvetic:1996uf} for examples). Therefore one can study properties of black holes with de Sitter asymptotic obtained within the effective low-energy models generated by the string theory. On the practical ground such models may be treated on the same foot as modified gravity models and should be addressed with the discussion from the previous sections. As string theory itself is not yet complete it is impossible to find an single modified gravity model providing a comprehensive low-energy description of the string theory. Therefore one should investigate various string-inspired models. 

\section{Conclusion} \label{concl}

In this paper we reviewed the properties of black hole like solutions in the scalar-tensor gravity. {Scalar-tensor theories play an important role in the xtended gravity layout. This key  role is defined by the fact that such models provide a way to separate a new gravitational degree of freedom at the level of the model action. This is a powerfull tool allowing one to trace the structure of a theory in a more detail. For instance in the $f(R)$ gravity models the new degree of freedom is encoded into a metric variable.}

Although the theories in the scope of this paper are very different, black hole solutions display several general properties:
\begin{itemize}[leftmargin=*,labelsep=5.8mm]
\item the model, describing the de Sitter expansion of the universe, naturally contains the asymptotically de Sitter solution;
\item the local phenomenology may be isolated from the cosmological expansion by means of the Vainshtein screening or the parameter fine tuning, which results in a flat asymptotic for the local solution.
\end{itemize}
Considering the cosmological and the  black hole solutions at the same time yields a more comprehensive view of the model and may result in useful constraints. For example, in the Brans-Dicke theory we know that we must account for the scalar field potential or the cosmological constant to obtain the cosmological expansion, therefore we must consider black holes in terms such models as well. In such a consideration different constraints may come into contradiction, thereby cancelling the model, as it occurred in \cite{Tretyakova:2015vaa,Tretyakova:2017lyg}.
	
However, such a versatile picture is rarely available. The unresolved questions about scalar-tensor theories in the scope of our paper are the following.
\begin{itemize}[leftmargin=*,labelsep=5.8mm]
\item There are no wormhole solutions for the BD+$\Lambda$ framework. 
\item Scalar-tensor models with the Gauss-Bonnet term should be explored in more detail to clarify the concordance between the local and the cosmological solution.
\item Gauss-Bonnet-Horndeski theory describes the late-time cosmology and inflation, but no analytic black hole solutions embedded in an expanding Universe is known. Such a solution could give a major insight into the model properties.
\item Quartic Horndeski square root Lagrangian is explored for black holes, however, we are lacking cosmological research.
\item {Horndeski models require additional research of black hole thermodynamic.}
\item It is not clear if most of the scalar-tensor black hole configurations actually take place during the matter collapse.
\item The no-hair argument should be considered for the theories with the auxiliary fields.
\end{itemize}

Unfortunately the asymptotic hevahior of the solution is not always known, even if the cosmology in the model is known to be adequate, like in \cite{Sotiriou:2014pfa}. This prevents the conformity assessment between the cosmological and astrophysical scale.

Despite all the issues mentioned above the scalar-tensor gravity remains one of the most fruitful extended gravity frameworks naturally emerging from modern theoretical and observational considerations. 

\acknowledgments{This work was supported by Russian Foundation for Basic Research via grant RFBR 16-02-00682.}

\externalbibliography{yes}
\bibliography{Biblio}

\begin{thebibliography}{-------}
\providecommand{\natexlab}[1]{#1}

\bibitem[Will(2014)]{Will:2014kxa}
Will, C.M.
\newblock {The Confrontation between General Relativity and Experiment}.
\newblock {\em Living Rev. Rel.} {\bf 2014}, {\em 17},~4,
  \href{http://xxx.lanl.gov/abs/1403.7377}{{\normalfont
  [arXiv:gr-qc/1403.7377]}}.

\bibitem[Hinshaw \em{et~al.}(2013)Hinshaw et~al.]{Hinshaw:2012aka}
Hinshaw, G.; others.
\newblock {Nine-Year Wilkinson Microwave Anisotropy Probe (WMAP) Observations:
  Cosmological Parameter Results}.
\newblock {\em Astrophys. J. Suppl.} {\bf 2013}, {\em 208},~19,
  \href{http://xxx.lanl.gov/abs/1212.5226}{{\normalfont
  [arXiv:astro-ph.CO/1212.5226]}}.

\bibitem[Ade \em{et~al.}(2016)Ade et~al.]{Ade:2015xua}
Ade, P.A.R.; others.
\newblock {Planck 2015 results. XIII. Cosmological parameters}.
\newblock {\em Astron. Astrophys.} {\bf 2016}, {\em 594},~A13,
  \href{http://xxx.lanl.gov/abs/1502.01589}{{\normalfont
  [arXiv:astro-ph.CO/1502.01589]}}.

\bibitem[Abbott \em{et~al.}(2016{\natexlab{a}})Abbott et~al.]{Abbott:2016blz}
Abbott, B.P.; others.
\newblock {Observation of Gravitational Waves from a Binary Black Hole Merger}.
\newblock {\em Phys. Rev. Lett.} {\bf 2016}, {\em 116},~061102,
  \href{http://xxx.lanl.gov/abs/1602.03837}{{\normalfont
  [arXiv:gr-qc/1602.03837]}}.

\bibitem[Abbott \em{et~al.}(2016{\natexlab{b}})Abbott et~al.]{Abbott:2016nmj}
Abbott, B.P.; others.
\newblock {GW151226: Observation of Gravitational Waves from a 22-Solar-Mass
  Binary Black Hole Coalescence}.
\newblock {\em Phys. Rev. Lett.} {\bf 2016}, {\em 116},~241103,
  \href{http://xxx.lanl.gov/abs/1606.04855}{{\normalfont
  [arXiv:gr-qc/1606.04855]}}.

\bibitem[Abbott \em{et~al.}(2017)Abbott et~al.]{Abbott:2017vtc}
Abbott, B.P.; others.
\newblock {GW170104: Observation of a 50-Solar-Mass Binary Black Hole
  Coalescence at Redshift 0.2}.
\newblock {\em Phys. Rev. Lett.} {\bf 2017}, {\em 118},~221101,
  \href{http://xxx.lanl.gov/abs/1706.01812}{{\normalfont
  [arXiv:gr-qc/1706.01812]}}.

\bibitem[Abbott \em{et~al.}(2016)Abbott et~al.]{TheLIGOScientific:2016src}
Abbott, B.P.; others.
\newblock {Tests of general relativity with GW150914}.
\newblock {\em Phys. Rev. Lett.} {\bf 2016}, {\em 116},~221101,
  \href{http://xxx.lanl.gov/abs/1602.03841}{{\normalfont
  [arXiv:gr-qc/1602.03841]}}.

\bibitem[Moffat(2016)]{Moffat:2016gkd}
Moffat, J.W.
\newblock {LIGO GW150914 and GW151226 gravitational wave detection and
  generalized gravitation theory (MOG)}.
\newblock {\em Phys. Lett.} {\bf 2016}, {\em B763},~427--433,
  \href{http://xxx.lanl.gov/abs/1603.05225}{{\normalfont
  [arXiv:gr-qc/1603.05225]}}.

\bibitem[Cayuso \em{et~al.}(2017)Cayuso, Ortiz, and Lehner]{Cayuso:2017iqc}
Cayuso, J.; Ortiz, N.; Lehner, L.
\newblock {Fixing extensions to General Relativity in the non-linear regime}.
\newblock {\em Phys. Rev.} {\bf 2017}, {\em D96},~084043,
  \href{http://xxx.lanl.gov/abs/1706.07421}{{\normalfont
  [arXiv:gr-qc/1706.07421]}}.

\bibitem[Abbott(2017{\natexlab{a}})]{PhysRevLett.119.141101}
Abbott, B.P.e.a.
\newblock GW170814: A Three-Detector Observation of Gravitational Waves from a
  Binary Black Hole Coalescence.
\newblock {\em Phys. Rev. Lett.} {\bf 2017}, {\em 119},~141101.

\bibitem[Abbott(2017{\natexlab{b}})]{PhysRevLett.119.161101}
Abbott, B.P.e.a.
\newblock GW170817: Observation of Gravitational Waves from a Binary Neutron
  Star Inspiral.
\newblock {\em Phys. Rev. Lett.} {\bf 2017}, {\em 119},~161101.

\bibitem[GBM \em{et~al.}(2017)GBM et~al.]{GBM:2017lvd}
GBM, F.; others.
\newblock {Multi-messenger Observations of a Binary Neutron Star Merger}.
\newblock {\em Astrophys. J.} {\bf 2017}, {\em 848},~L12,
  \href{http://xxx.lanl.gov/abs/1710.05833}{{\normalfont
  [arXiv:astro-ph.HE/1710.05833]}}.

\bibitem[Fish \em{et~al.}(2016)Fish, Akiyama, Bouman, Chael, Johnson, Doeleman,
  Blackburn, Wardle, and Freeman]{Fish_2016}
Fish, V.; Akiyama, K.; Bouman, K.; Chael, A.; Johnson, M.; Doeleman, S.;
  Blackburn, L.; Wardle, J.; Freeman, W.
\newblock Observing—and Imaging—Active Galactic Nuclei with the Event
  Horizon Telescope.
\newblock {\em Galaxies} {\bf 2016}, {\em 4},~54.

\bibitem[Ortiz-León \em{et~al.}(2016)Ortiz-León et~al.]{Ortiz-Leon:2016cch}
Ortiz-León, G.N.; others.
\newblock {The Intrinsic Shape of Sagittarius A* at 3.5-mm Wavelength}.
\newblock {\em Astrophys. J.} {\bf 2016}, {\em 824},~40,
  \href{http://xxx.lanl.gov/abs/1601.06571}{{\normalfont
  [arXiv:astro-ph.GA/1601.06571]}}.

\bibitem[Berti \em{et~al.}(2015)Berti et~al.]{Berti:2015itd}
Berti, E.; others.
\newblock {Testing General Relativity with Present and Future Astrophysical
  Observations}.
\newblock {\em Class. Quant. Grav.} {\bf 2015}, {\em 32},~243001,
  \href{http://xxx.lanl.gov/abs/1501.07274}{{\normalfont
  [arXiv:gr-qc/1501.07274]}}.

\bibitem[Capozziello and De~Laurentis(2011)]{Capozziello:2011et}
Capozziello, S.; De~Laurentis, M.
\newblock {Extended Theories of Gravity}.
\newblock {\em Phys. Rept.} {\bf 2011}, {\em 509},~167--321,
  \href{http://xxx.lanl.gov/abs/1108.6266}{{\normalfont
  [arXiv:gr-qc/1108.6266]}}.

\bibitem[Bertotti \em{et~al.}(2003)Bertotti, Iess, and
  Tortora]{Bertotti:2003rm}
Bertotti, B.; Iess, L.; Tortora, P.
\newblock {A test of general relativity using radio links with the Cassini
  spacecraft}.
\newblock {\em Nature} {\bf 2003}, {\em 425},~374--376.

\bibitem[Clowe \em{et~al.}(2006)Clowe, Bradac, Gonzalez, Markevitch, Randall,
  Jones, and Zaritsky]{Clowe:2006eq}
Clowe, D.; Bradac, M.; Gonzalez, A.H.; Markevitch, M.; Randall, S.W.; Jones,
  C.; Zaritsky, D.
\newblock {A direct empirical proof of the existence of dark matter}.
\newblock {\em Astrophys. J.} {\bf 2006}, {\em 648},~L109--L113,
  \href{http://xxx.lanl.gov/abs/astro-ph/0608407}{{\normalfont
  [arXiv:astro-ph/astro-ph/0608407]}}.

\bibitem[Boran \em{et~al.}(2017)Boran, Desai, Kahya, and
  Woodard]{Boran:2017rdn}
Boran, S.; Desai, S.; Kahya, E.; Woodard, R.
\newblock {GW170817 Falsifies Dark Matter Emulators} {\bf 2017}.
\newblock  \href{http://xxx.lanl.gov/abs/1710.06168}{{\normalfont
  [arXiv:astro-ph.HE/1710.06168]}}.

\bibitem[Blumenthal \em{et~al.}(1984)Blumenthal, Faber, Primack, and
  Rees]{Blumenthal:1984bp}
Blumenthal, G.R.; Faber, S.M.; Primack, J.R.; Rees, M.J.
\newblock {Formation of Galaxies and Large Scale Structure with Cold Dark
  Matter}.
\newblock {\em Nature} {\bf 1984}, {\em 311},~517--525.

\bibitem[Davis \em{et~al.}(1985)Davis, Efstathiou, Frenk, and
  White]{Davis:1985rj}
Davis, M.; Efstathiou, G.; Frenk, C.S.; White, S.D.M.
\newblock {The Evolution of Large Scale Structure in a Universe Dominated by
  Cold Dark Matter}.
\newblock {\em Astrophys. J.} {\bf 1985}, {\em 292},~371--394.

\bibitem[Capozziello \em{et~al.}(2004)Capozziello, Cardone, Carloni, and
  Troisi]{Capozziello:2004us}
Capozziello, S.; Cardone, V.F.; Carloni, S.; Troisi, A.
\newblock {Can higher order curvature theories explain rotation curves of
  galaxies?}
\newblock {\em Phys. Lett.} {\bf 2004}, {\em A326},~292--296,
  \href{http://xxx.lanl.gov/abs/gr-qc/0404114}{{\normalfont
  [arXiv:gr-qc/gr-qc/0404114]}}.

\bibitem[Capozziello \em{et~al.}(2006)Capozziello, Cardone, and
  Troisi]{Capozziello:2006uv}
Capozziello, S.; Cardone, V.F.; Troisi, A.
\newblock {Dark energy and dark matter as curvature effects}.
\newblock {\em JCAP} {\bf 2006}, {\em 0608},~001,
  \href{http://xxx.lanl.gov/abs/astro-ph/0602349}{{\normalfont
  [arXiv:astro-ph/astro-ph/0602349]}}.

\bibitem[Capozziello \em{et~al.}(2007)Capozziello, Cardone, and
  Troisi]{Capozziello:2006ph}
Capozziello, S.; Cardone, V.F.; Troisi, A.
\newblock {Low surface brightness galaxies rotation curves in the low energy
  limit of r**n gravity: no need for dark matter?}
\newblock {\em Mon. Not. Roy. Astron. Soc.} {\bf 2007}, {\em 375},~1423--1440,
  \href{http://xxx.lanl.gov/abs/astro-ph/0603522}{{\normalfont
  [arXiv:astro-ph/astro-ph/0603522]}}.

\bibitem[Zel'dovich \em{et~al.}(1968)Zel'dovich, Krasinski, and
  Zeldovich]{Zeldovich:1968ehl}
Zel'dovich, {\relax Ya}.B.; Krasinski, A.; Zeldovich, {\relax Ya}.B.
\newblock {The Cosmological constant and the theory of elementary particles}.
\newblock {\em Sov. Phys. Usp.} {\bf 1968}, {\em 11},~381--393.
\newblock [Usp. Fiz. Nauk95,209(1968)].

\bibitem[Weinberg(1989)]{Weinberg:1988cp}
Weinberg, S.
\newblock {The Cosmological Constant Problem}.
\newblock {\em Rev. Mod. Phys.} {\bf 1989}, {\em 61},~1--23.

\bibitem[Linde(2008)]{Linde:2007fr}
Linde, A.D.
\newblock {Inflationary Cosmology}.
\newblock {\em Lect. Notes Phys.} {\bf 2008}, {\em 738},~1--54,
  \href{http://xxx.lanl.gov/abs/0705.0164}{{\normalfont
  [arXiv:hep-th/0705.0164]}}.

\bibitem[Senatore(2017)]{Senatore:2016aui}
Senatore, L.
\newblock {Lectures on Inflation}.
\newblock  {Proceedings, Theoretical Advanced Study Institute in Elementary
  Particle Physics: New Frontiers in Fields and Strings (TASI 2015): Boulder,
  CO, USA, June 1-26, 2015},  2017, pp. 447--543,
  \href{http://xxx.lanl.gov/abs/1609.00716}{{\normalfont
  [arXiv:hep-th/1609.00716]}}.

\bibitem[Starobinsky(1980)]{Starobinsky:1980te}
Starobinsky, A.A.
\newblock {A New Type of Isotropic Cosmological Models Without Singularity}.
\newblock {\em Phys. Lett.} {\bf 1980}, {\em 91B},~99--102.

\bibitem[Linde(1983)]{Linde:1983gd}
Linde, A.D.
\newblock {Chaotic Inflation}.
\newblock {\em Phys. Lett.} {\bf 1983}, {\em 129B},~177--181.

\bibitem[De~Felice and Tsujikawa(2010)]{DeFelice:2010aj}
De~Felice, A.; Tsujikawa, S.
\newblock {f(R) theories}.
\newblock {\em Living Rev. Rel.} {\bf 2010}, {\em 13},~3,
  \href{http://xxx.lanl.gov/abs/1002.4928}{{\normalfont
  [arXiv:gr-qc/1002.4928]}}.

\bibitem[Nojiri \em{et~al.}(2017)Nojiri, Odintsov, and
  Oikonomou]{Nojiri:2017ncd}
Nojiri, S.; Odintsov, S.D.; Oikonomou, V.K.
\newblock {Modified Gravity Theories on a Nutshell: Inflation, Bounce and
  Late-time Evolution}.
\newblock {\em Phys. Rept.} {\bf 2017}, {\em 692},~1--104,
  \href{http://xxx.lanl.gov/abs/1705.11098}{{\normalfont
  [arXiv:gr-qc/1705.11098]}}.

\bibitem[Nojiri and Odintsov(2011)]{Nojiri:2010wj}
Nojiri, S.; Odintsov, S.D.
\newblock {Unified cosmic history in modified gravity: from F(R) theory to
  Lorentz non-invariant models}.
\newblock {\em Phys. Rept.} {\bf 2011}, {\em 505},~59--144,
  \href{http://xxx.lanl.gov/abs/1011.0544}{{\normalfont
  [arXiv:gr-qc/1011.0544]}}.

\bibitem[Nojiri and Odintsov(2006)]{Nojiri:2006ri}
Nojiri, S.; Odintsov, S.D.
\newblock {Introduction to modified gravity and gravitational alternative for
  dark energy}.
\newblock {\em eConf} {\bf 2006}, {\em C0602061},~06,
  \href{http://xxx.lanl.gov/abs/hep-th/0601213}{{\normalfont
  [arXiv:hep-th/hep-th/0601213]}}.
\newblock [Int. J. Geom. Meth. Mod. Phys.4,115(2007)].

\bibitem[Charmousis \em{et~al.}(2012)Charmousis, Copeland, Padilla, and
  Saffin]{Charmousis:2011bf}
Charmousis, C.; Copeland, E.J.; Padilla, A.; Saffin, P.M.
\newblock {General second order scalar-tensor theory, self tuning, and the Fab
  Four}.
\newblock {\em Phys. Rev. Lett.} {\bf 2012}, {\em 108},~051101,
  \href{http://xxx.lanl.gov/abs/1106.2000}{{\normalfont
  [arXiv:hep-th/1106.2000]}}.

\bibitem[Momeni \em{et~al.}(2016)Momeni, Houndjo, Güdekli, Rodrigues,
  Alvarenga, and Myrzakulov]{Momeni:2014uwa}
Momeni, D.; Houndjo, M.J.S.; Güdekli, E.; Rodrigues, M.E.; Alvarenga, F.G.;
  Myrzakulov, R.
\newblock {Spherically symmetric solutions of light Galileon}.
\newblock {\em Int. J. Theor. Phys.} {\bf 2016}, {\em 55},~1211--1221,
  \href{http://xxx.lanl.gov/abs/1412.4672}{{\normalfont
  [arXiv:gr-qc/1412.4672]}}.

\bibitem[Brans and Dicke(1961)]{Brans:1961sx}
Brans, C.; Dicke, R.H.
\newblock {Mach's principle and a relativistic theory of gravitation}.
\newblock {\em Phys. Rev.} {\bf 1961}, {\em 124},~925--935.

\bibitem[de~la Cruz-Dombriz and Saez-Gomez(2012)]{delaCruzDombriz:2012xy}
de~la Cruz-Dombriz, A.; Saez-Gomez, D.
\newblock {Black holes, cosmological solutions, future singularities, and their
  thermodynamical properties in modified gravity theories}.
\newblock {\em Entropy} {\bf 2012}, {\em 14},~1717--1770,
  \href{http://xxx.lanl.gov/abs/1207.2663}{{\normalfont
  [arXiv:gr-qc/1207.2663]}}.

\bibitem[Bergmann(1968)]{Bergmann:1968ve}
Bergmann, P.G.
\newblock {Comments on the scalar tensor theory}.
\newblock {\em Int. J. Theor. Phys.} {\bf 1968}, {\em 1},~25--36.

\bibitem[Wagoner(1970)]{Wagoner:1970vr}
Wagoner, R.V.
\newblock {Scalar tensor theory and gravitational waves}.
\newblock {\em Phys. Rev.} {\bf 1970}, {\em D1},~3209--3216.

\bibitem[Nordtvedt(1970)]{Nordtvedt:1970uv}
Nordtvedt, Jr., K.
\newblock {PostNewtonian metric for a general class of scalar tensor
  gravitational theories and observational consequences}.
\newblock {\em Astrophys. J.} {\bf 1970}, {\em 161},~1059--1067.

\bibitem[Clifton \em{et~al.}(2012)Clifton, Ferreira, Padilla, and
  Skordis]{Clifton:2011jh}
Clifton, T.; Ferreira, P.G.; Padilla, A.; Skordis, C.
\newblock {Modified Gravity and Cosmology}.
\newblock {\em Phys. Rept.} {\bf 2012}, {\em 513},~1--189,
  \href{http://xxx.lanl.gov/abs/1106.2476}{{\normalfont
  [arXiv:astro-ph.CO/1106.2476]}}.

\bibitem[Hohmann \em{et~al.}(2013)Hohmann, Jarv, Kuusk, and
  Randla]{Hohmann:2013rba}
Hohmann, M.; Jarv, L.; Kuusk, P.; Randla, E.
\newblock {Post-Newtonian parameters $\gamma$ and $\beta$ of scalar-tensor
  gravity with a general potential}.
\newblock {\em Phys. Rev.} {\bf 2013}, {\em D88},~084054,
  \href{http://xxx.lanl.gov/abs/1309.0031}{{\normalfont
  [arXiv:gr-qc/1309.0031]}}.
\newblock [Erratum: Phys. Rev.D89,no.6,069901(2014)].

\bibitem[Deng and Xie(2016)]{Deng:2016moh}
Deng, X.M.; Xie, Y.
\newblock {Solar System tests of a scalar-tensor gravity with a general
  potential: Insensitivity of light deflection and Cassini tracking}.
\newblock {\em Phys. Rev.} {\bf 2016}, {\em D93},~044013.

\bibitem[Tretyakova \em{et~al.}(2012)Tretyakova, Shatskiy, Novikov, and
  Alexeyev]{Tretyakova2011ch}
Tretyakova, D.; Shatskiy, A.; Novikov, I.; Alexeyev, S.
\newblock {Non-singular Brans-Dicke cosmology with cosmological constant}.
\newblock {\em Phys.Rev.} {\bf 2012}, {\em D85},~124059,
  \href{http://xxx.lanl.gov/abs/1112.3770}{{\normalfont
  [arXiv:gr-qc/1112.3770]}}.

\bibitem[Damour and Esposito-Farese(1993)]{Damour:1993hw}
Damour, T.; Esposito-Farese, G.
\newblock {Nonperturbative strong field effects in tensor - scalar theories of
  gravitation}.
\newblock {\em Phys. Rev. Lett.} {\bf 1993}, {\em 70},~2220--2223.

\bibitem[Damour and Esposito-Farese(1996)]{Damour:1996ke}
Damour, T.; Esposito-Farese, G.
\newblock {Tensor - scalar gravity and binary pulsar experiments}.
\newblock {\em Phys. Rev.} {\bf 1996}, {\em D54},~1474--1491,
  \href{http://xxx.lanl.gov/abs/gr-qc/9602056}{{\normalfont
  [arXiv:gr-qc/gr-qc/9602056]}}.

\bibitem[Shibata \em{et~al.}(1994)Shibata, Nakao, and Nakamura]{Shibata:1994qd}
Shibata, M.; Nakao, K.i.; Nakamura, T.
\newblock {Scalar type gravitational wave emission from gravitational collapse
  in Brans-Dicke theory: Detectability by a laser interferometer}.
\newblock {\em Phys. Rev.} {\bf 1994}, {\em D50},~7304--7317.

\bibitem[Harada \em{et~al.}(1997)Harada, Chiba, Nakao, and
  Nakamura]{Harada:1996wt}
Harada, T.; Chiba, T.; Nakao, K.i.; Nakamura, T.
\newblock {Scalar gravitational wave from Oppenheimer-Snyder collapse in scalar
  - tensor theories of gravity}.
\newblock {\em Phys. Rev.} {\bf 1997}, {\em D55},~2024--2037,
  \href{http://xxx.lanl.gov/abs/gr-qc/9611031}{{\normalfont
  [arXiv:gr-qc/gr-qc/9611031]}}.

\bibitem[Novak(1998)]{Novak:1997hw}
Novak, J.
\newblock {Spherical neutron star collapse in tensor - scalar theory of
  gravity}.
\newblock {\em Phys. Rev.} {\bf 1998}, {\em D57},~4789--4801,
  \href{http://xxx.lanl.gov/abs/gr-qc/9707041}{{\normalfont
  [arXiv:gr-qc/gr-qc/9707041]}}.

\bibitem[Faraoni \em{et~al.}(2015)Faraoni, Lapierre-Léonard, and
  Prain]{Faraoni:2015saa}
Faraoni, V.; Lapierre-Léonard, M.; Prain, A.
\newblock {Turnaround radius in an accelerated universe with quasi-local mass}.
\newblock {\em JCAP} {\bf 2015}, {\em 1510},~013,
  \href{http://xxx.lanl.gov/abs/1508.01725}{{\normalfont
  [arXiv:gr-qc/1508.01725]}}.

\bibitem[Faraoni(2016)]{Faraoni:2015zqa}
Faraoni, V.
\newblock {Turnaround radius in modified gravity}.
\newblock {\em Phys. Dark Univ.} {\bf 2016}, {\em 11},~11--15,
  \href{http://xxx.lanl.gov/abs/1508.00475}{{\normalfont
  [arXiv:gr-qc/1508.00475]}}.

\bibitem[Pavlidou and Tomaras(2014)]{Pavlidou:2013zha}
Pavlidou, V.; Tomaras, T.N.
\newblock {Where the world stands still: turnaround as a strong test of
  $\Lambda$CDM cosmology}.
\newblock {\em JCAP} {\bf 2014}, {\em 1409},~020,
  \href{http://xxx.lanl.gov/abs/1310.1920}{{\normalfont
  [arXiv:astro-ph.CO/1310.1920]}}.

\bibitem[Pavlidou \em{et~al.}(2014)Pavlidou, Tetradis, and
  Tomaras]{Pavlidou:2014aia}
Pavlidou, V.; Tetradis, N.; Tomaras, T.N.
\newblock {Constraining Dark Energy through the Stability of Cosmic
  Structures}.
\newblock {\em JCAP} {\bf 2014}, {\em 1405},~017,
  \href{http://xxx.lanl.gov/abs/1401.3742}{{\normalfont
  [arXiv:astro-ph.CO/1401.3742]}}.

\bibitem[Tanoglidis \em{et~al.}(2015)Tanoglidis, Pavlidou, and
  Tomaras]{Tanoglidis:2014lea}
Tanoglidis, D.; Pavlidou, V.; Tomaras, T.
\newblock {Testing LambdaCDM cosmology at turnaround: where to look for
  violations of the bound?}
\newblock {\em JCAP} {\bf 2015}, {\em 1512},~060,
  \href{http://xxx.lanl.gov/abs/1412.6671}{{\normalfont
  [arXiv:astro-ph.CO/1412.6671]}}.

\bibitem[Tretyakova \em{et~al.}(2012)Tretyakova, Shatskiy, Novikov, and
  Alexeyev]{Tretyakova:2011ch}
Tretyakova, D.A.; Shatskiy, A.A.; Novikov, I.D.; Alexeyev, S.
\newblock {Non-singular Brans-Dicke cosmology with cosmological constant}.
\newblock {\em Phys. Rev.} {\bf 2012}, {\em D85},~124059,
  \href{http://xxx.lanl.gov/abs/1112.3770}{{\normalfont
  [arXiv:gr-qc/1112.3770]}}.

\bibitem[Damour and Nordtvedt(1993)]{Damour:1992kf}
Damour, T.; Nordtvedt, K.
\newblock {General relativity as a cosmological attractor of tensor scalar
  theories}.
\newblock {\em Phys. Rev. Lett.} {\bf 1993}, {\em 70},~2217--2219.

\bibitem[Elizalde \em{et~al.}(2004)Elizalde, Nojiri, and
  Odintsov]{Elizalde:2004mq}
Elizalde, E.; Nojiri, S.; Odintsov, S.D.
\newblock {Late-time cosmology in (phantom) scalar-tensor theory: Dark energy
  and the cosmic speed-up}.
\newblock {\em Phys. Rev.} {\bf 2004}, {\em D70},~043539,
  \href{http://xxx.lanl.gov/abs/hep-th/0405034}{{\normalfont
  [arXiv:hep-th/hep-th/0405034]}}.

\bibitem[Ratra and Peebles(1988)]{Ratra:1987rm}
Ratra, B.; Peebles, P.J.E.
\newblock {Cosmological Consequences of a Rolling Homogeneous Scalar Field}.
\newblock {\em Phys. Rev.} {\bf 1988}, {\em D37},~3406.

\bibitem[Demianski \em{et~al.}(2008)Demianski, Piedipalumbo, Rubano, and
  Scudellaro]{Demianski:2007mz}
Demianski, M.; Piedipalumbo, E.; Rubano, C.; Scudellaro, P.
\newblock {Cosmological models in scalar tensor theories of gravity and
  observations: A class of general solutions}.
\newblock {\em Astron. Astrophys.} {\bf 2008}, {\em 481},~279--294,
  \href{http://xxx.lanl.gov/abs/0711.1043}{{\normalfont
  [arXiv:astro-ph/0711.1043]}}.

\bibitem[NAGATA(2011)]{doi:10.1142/S2010194511000250}
NAGATA, R.
\newblock CONSTRAINTS ON SCALAR-TENSOR COSMOLOGY FROM WMAP DATA.
\newblock {\em International Journal of Modern Physics: Conference Series} {\bf
  2011}, {\em 01},~183--188,
  \href{http://xxx.lanl.gov/abs/http://www.worldscientific.com/doi/pdf/10.1142/S2010194511000250}{{\normalfont
  [http://www.worldscientific.com/doi/pdf/10.1142/S2010194511000250]}}.

\bibitem[Alonso \em{et~al.}(2017)Alonso, Bellini, Ferreira, and
  Zumalac\'arregui]{PhysRevD.95.063502}
Alonso, D.; Bellini, E.; Ferreira, P.G.; Zumalac\'arregui, M.
\newblock Observational future of cosmological scalar-tensor theories.
\newblock {\em Phys. Rev. D} {\bf 2017}, {\em 95},~063502.

\bibitem[Babichev and Deffayet(2013)]{Babichev:2013usa}
Babichev, E.; Deffayet, C.
\newblock {An introduction to the Vainshtein mechanism}.
\newblock {\em Class. Quant. Grav.} {\bf 2013}, {\em 30},~184001,
  \href{http://xxx.lanl.gov/abs/1304.7240}{{\normalfont
  [arXiv:gr-qc/1304.7240]}}.

\bibitem[Bellini and Sawicki(2014)]{Bellini:2014fua}
Bellini, E.; Sawicki, I.
\newblock {Maximal freedom at minimum cost: linear large-scale structure in
  general modifications of gravity}.
\newblock {\em JCAP} {\bf 2014}, {\em 1407},~050,
  \href{http://xxx.lanl.gov/abs/1404.3713}{{\normalfont
  [arXiv:astro-ph.CO/1404.3713]}}.

\bibitem[Magnano and Sokolowski(1994)]{Magnano:1993bd}
Magnano, G.; Sokolowski, L.M.
\newblock {On physical equivalence between nonlinear gravity theories and a
  general relativistic selfgravitating scalar field}.
\newblock {\em Phys. Rev.} {\bf 1994}, {\em D50},~5039--5059,
  \href{http://xxx.lanl.gov/abs/gr-qc/9312008}{{\normalfont
  [arXiv:gr-qc/gr-qc/9312008]}}.

\bibitem[Faraoni and Gunzig(1999)]{Faraoni:1999hp}
Faraoni, V.; Gunzig, E.
\newblock {Einstein frame or Jordan frame?}
\newblock {\em Int. J. Theor. Phys.} {\bf 1999}, {\em 38},~217--225,
  \href{http://xxx.lanl.gov/abs/astro-ph/9910176}{{\normalfont
  [arXiv:astro-ph/astro-ph/9910176]}}.

\bibitem[Postma and Volponi(2014)]{Postma:2014vaa}
Postma, M.; Volponi, M.
\newblock {Equivalence of the Einstein and Jordan frames}.
\newblock {\em Phys. Rev.} {\bf 2014}, {\em D90},~103516,
  \href{http://xxx.lanl.gov/abs/1407.6874}{{\normalfont
  [arXiv:astro-ph.CO/1407.6874]}}.

\bibitem[Catena \em{et~al.}(2007)Catena, Pietroni, and
  Scarabello]{Catena:2006bd}
Catena, R.; Pietroni, M.; Scarabello, L.
\newblock {Einstein and Jordan reconciled: a frame-invariant approach to
  scalar-tensor cosmology}.
\newblock {\em Phys. Rev.} {\bf 2007}, {\em D76},~084039,
  \href{http://xxx.lanl.gov/abs/astro-ph/0604492}{{\normalfont
  [arXiv:astro-ph/astro-ph/0604492]}}.

\bibitem[Faraoni and Nadeau(2007)]{Faraoni:2006fx}
Faraoni, V.; Nadeau, S.
\newblock {The (pseudo)issue of the conformal frame revisited}.
\newblock {\em Phys. Rev.} {\bf 2007}, {\em D75},~023501,
  \href{http://xxx.lanl.gov/abs/gr-qc/0612075}{{\normalfont
  [arXiv:gr-qc/gr-qc/0612075]}}.

\bibitem[Salgado(2006)]{Salgado:2005hx}
Salgado, M.
\newblock {The Cauchy problem of scalar tensor theories of gravity}.
\newblock {\em Class. Quant. Grav.} {\bf 2006}, {\em 23},~4719--4742,
  \href{http://xxx.lanl.gov/abs/gr-qc/0509001}{{\normalfont
  [arXiv:gr-qc/gr-qc/0509001]}}.

\bibitem[Bahamonde \em{et~al.}(2016)Bahamonde, Odintsov, Oikonomou, and
  Wright]{Bahamonde:2016wmz}
Bahamonde, S.; Odintsov, S.D.; Oikonomou, V.K.; Wright, M.
\newblock {Correspondence of $F(R)$ Gravity Singularities in Jordan and
  Einstein Frames}.
\newblock {\em Annals Phys.} {\bf 2016}, {\em 373},~96--114,
  \href{http://xxx.lanl.gov/abs/1603.05113}{{\normalfont
  [arXiv:gr-qc/1603.05113]}}.

\bibitem[Bahamonde \em{et~al.}(2017)Bahamonde, Odintsov, Oikonomou, and
  Tretyakov]{Bahamonde:2017kbs}
Bahamonde, S.; Odintsov, S.D.; Oikonomou, V.K.; Tretyakov, P.V.
\newblock {Deceleration versus acceleration universe in different frames of
  $F(R)$ gravity}.
\newblock {\em Phys. Lett.} {\bf 2017}, {\em B766},~225--230,
  \href{http://xxx.lanl.gov/abs/1701.02381}{{\normalfont
  [arXiv:gr-qc/1701.02381]}}.

\bibitem[Calmet and Yang(2013)]{Calmet:2012eq}
Calmet, X.; Yang, T.C.
\newblock {Frame Transformations of Gravitational Theories}.
\newblock {\em Int. J. Mod. Phys.} {\bf 2013}, {\em A28},~1350042,
  \href{http://xxx.lanl.gov/abs/1211.4217}{{\normalfont
  [arXiv:gr-qc/1211.4217]}}.

\bibitem[Pandey and Banerjee(2017)]{Pandey:2016unk}
Pandey, S.; Banerjee, N.
\newblock {Equivalence of Jordan and Einstein frames at the quantum level}.
\newblock {\em Eur. Phys. J. Plus} {\bf 2017}, {\em 132},~107,
  \href{http://xxx.lanl.gov/abs/1610.00584}{{\normalfont
  [arXiv:gr-qc/1610.00584]}}.

\bibitem[Kamenshchik and Steinwachs(2015)]{Kamenshchik:2014waa}
Kamenshchik, A.{\relax Yu}.; Steinwachs, C.F.
\newblock {Question of quantum equivalence between Jordan frame and Einstein
  frame}.
\newblock {\em Phys. Rev.} {\bf 2015}, {\em D91},~084033,
  \href{http://xxx.lanl.gov/abs/1408.5769}{{\normalfont
  [arXiv:gr-qc/1408.5769]}}.

\bibitem[Kottler(1918)]{kottler}
Kottler, F.
\newblock The physical basis of Einstein's theory of gravitation.
\newblock {\em Ann.\ Phys.\ (Leipzig)} {\bf 1918}, p. 401.

\bibitem[Ashtekar and Magnon(1984)]{Ashtekar:1984zz}
Ashtekar, A.; Magnon, A.
\newblock {Asymptotically anti-de Sitter space-times}.
\newblock {\em Class. Quant. Grav.} {\bf 1984}, {\em 1},~L39--L44.

\bibitem[Kodama and Ishibashi(2003)]{Kodama:2003jz}
Kodama, H.; Ishibashi, A.
\newblock {A Master equation for gravitational perturbations of maximally
  symmetric black holes in higher dimensions}.
\newblock {\em Prog. Theor. Phys.} {\bf 2003}, {\em 110},~701--722,
  \href{http://xxx.lanl.gov/abs/hep-th/0305147}{{\normalfont
  [arXiv:hep-th/hep-th/0305147]}}.

\bibitem[Hawking and Page(1983)]{Hawking:1982dh}
Hawking, S.W.; Page, D.N.
\newblock {Thermodynamics of Black Holes in anti-De Sitter Space}.
\newblock {\em Commun. Math. Phys.} {\bf 1983}, {\em 87},~577.

\bibitem[Sotiriou(2015)]{Sotiriou:2015pka}
Sotiriou, T.P.
\newblock {Black Holes and Scalar Fields}.
\newblock {\em Class. Quant. Grav.} {\bf 2015}, {\em 32},~214002,
  \href{http://xxx.lanl.gov/abs/1505.00248}{{\normalfont
  [arXiv:gr-qc/1505.00248]}}.

\bibitem[Hawking(1972)]{Hawking:1971vc}
Hawking, S.W.
\newblock {Black holes in general relativity}.
\newblock {\em Commun. Math. Phys.} {\bf 1972}, {\em 25},~152--166.

\bibitem[Torii \em{et~al.}(2001)Torii, Maeda, and Narita]{PhysRevD.64.044007}
Torii, T.; Maeda, K.; Narita, M.
\newblock Scalar hair on the black hole in asymptotically anti\char21{}de
  Sitter spacetime.
\newblock {\em Phys. Rev. D} {\bf 2001}, {\em 64},~044007.

\bibitem[Sotiriou and Zhou(2014{\natexlab{a}})]{Sotiriou:2013qea}
Sotiriou, T.P.; Zhou, S.Y.
\newblock {Black hole hair in generalized scalar-tensor gravity}.
\newblock {\em Phys. Rev. Lett.} {\bf 2014}, {\em 112},~251102,
  \href{http://xxx.lanl.gov/abs/1312.3622}{{\normalfont
  [arXiv:gr-qc/1312.3622]}}.

\bibitem[Sotiriou and Zhou(2014{\natexlab{b}})]{Sotiriou:2014pfa}
Sotiriou, T.P.; Zhou, S.Y.
\newblock {Black hole hair in generalized scalar-tensor gravity: An explicit
  example}.
\newblock {\em Phys. Rev.} {\bf 2014}, {\em D90},~124063,
  \href{http://xxx.lanl.gov/abs/1408.1698}{{\normalfont
  [arXiv:gr-qc/1408.1698]}}.

\bibitem[Brans(1962)]{Brans:1962zz}
Brans, C.H.
\newblock {Mach's Principle and a Relativistic Theory of Gravitation. II}.
\newblock {\em Phys. Rev.} {\bf 1962}, {\em 125},~2194--2201.

\bibitem[Bhadra and Nandi(2001)]{Bhadra:2001fx}
Bhadra, A.; Nandi, K.K.
\newblock {On the equivalence of the Buchdahl and the Janis-Newman-Winnicour
  solutions}.
\newblock {\em Int. J. Mod. Phys.} {\bf 2001}, {\em A16},~4543--4545.

\bibitem[Bhadra and Sarkar(2005)]{Bhadra:2005mc}
Bhadra, A.; Sarkar, K.
\newblock {On static spherically symmetric solutions of the vacuum Brans-Dicke
  theory}.
\newblock {\em Gen. Rel. Grav.} {\bf 2005}, {\em 37},~2189--2199,
  \href{http://xxx.lanl.gov/abs/gr-qc/0505141}{{\normalfont
  [arXiv:gr-qc/gr-qc/0505141]}}.

\bibitem[Wyman(1981)]{Wyman:1981bd}
Wyman, M.
\newblock {Static Spherically Symmetric Scalar Fields in General Relativity}.
\newblock {\em Phys. Rev.} {\bf 1981}, {\em D24},~839--841.

\bibitem[Buchdahl(1959)]{Buchdahl:1959nk}
Buchdahl, H.A.
\newblock {Reciprocal Static Metrics and Scalar Fields in the General Theory of
  Relativity}.
\newblock {\em Phys. Rev.} {\bf 1959}, {\em 115},~1325--1328.

\bibitem[Agnese and La~Camera(1995)]{Agnese:1995kd}
Agnese, A.G.; La~Camera, M.
\newblock {Wormholes in the Brans-Dicke theory of gravitation}.
\newblock {\em Phys. Rev.} {\bf 1995}, {\em D51},~2011--2013.

\bibitem[Faraoni \em{et~al.}(2016)Faraoni, Hammad, and
  Belknap-Keet]{Faraoni:2016ozb}
Faraoni, V.; Hammad, F.; Belknap-Keet, S.D.
\newblock {Revisiting the Brans solutions of scalar-tensor gravity}.
\newblock {\em Phys. Rev.} {\bf 2016}, {\em D94},~104019,
  \href{http://xxx.lanl.gov/abs/1609.02783}{{\normalfont
  [arXiv:gr-qc/1609.02783]}}.

\bibitem[Bhattacharya \em{et~al.}(2015)Bhattacharya, Dialektopoulos, Romano,
  and Tomaras]{Bhattacharya:2015iha}
Bhattacharya, S.; Dialektopoulos, K.F.; Romano, A.E.; Tomaras, T.N.
\newblock {Brans-Dicke Theory with $\Lambda>0$: Black Holes and Large Scale
  Structures}.
\newblock {\em Phys. Rev. Lett.} {\bf 2015}, {\em 115},~181104,
  \href{http://xxx.lanl.gov/abs/1505.02375}{{\normalfont
  [arXiv:gr-qc/1505.02375]}}.

\bibitem[Kim(1997)]{Kim:1996fm}
Kim, H.
\newblock {Thermodynamics of black holes in Brans-Dicke gravity}.
\newblock {\em Nuovo Cim.} {\bf 1997}, {\em B112},~329--338,
  \href{http://xxx.lanl.gov/abs/gr-qc/9706044}{{\normalfont
  [arXiv:gr-qc/gr-qc/9706044]}}.

\bibitem[Faraoni(2010)]{Faraoni:2010yi}
Faraoni, V.
\newblock {Black hole entropy in scalar-tensor and f(R) gravity: An Overview}.
\newblock {\em Entropy} {\bf 2010}, {\em 12},~1246,
  \href{http://xxx.lanl.gov/abs/1005.2327}{{\normalfont
  [arXiv:gr-qc/1005.2327]}}.

\bibitem[Kang(1996)]{Kang:1996rj}
Kang, G.
\newblock {On black hole area in Brans-Dicke theory}.
\newblock {\em Phys. Rev.} {\bf 1996}, {\em D54},~7483--7489,
  \href{http://xxx.lanl.gov/abs/gr-qc/9606020}{{\normalfont
  [arXiv:gr-qc/gr-qc/9606020]}}.

\bibitem[Zaslavskii(2002)]{Zaslavskii:2002zv}
Zaslavskii, O.B.
\newblock {Thermodynamics of black holes with an infinite effective area of a
  horizon}.
\newblock {\em Class. Quant. Grav.} {\bf 2002}, {\em 19},~3783--3798,
  \href{http://xxx.lanl.gov/abs/gr-qc/0206018}{{\normalfont
  [arXiv:gr-qc/gr-qc/0206018]}}.

\bibitem[Nandi \em{et~al.}(2001)Nandi, Bhadra, Alsing, and Nayak]{Nandi:2000gt}
Nandi, K.K.; Bhadra, A.; Alsing, P.M.; Nayak, T.B.
\newblock {Tidal forces in cold black hole space-times}.
\newblock {\em Int. J. Mod. Phys.} {\bf 2001}, {\em D10},~529--538,
  \href{http://xxx.lanl.gov/abs/gr-qc/0008025}{{\normalfont
  [arXiv:gr-qc/gr-qc/0008025]}}.

\bibitem[Bronnikov \em{et~al.}(1998)Bronnikov, Clement, Constantinidis, and
  Fabris]{Bronnikov:1998hm}
Bronnikov, K.A.; Clement, G.; Constantinidis, C.P.; Fabris, J.C.
\newblock {Cold scalar tensor black holes: Causal structure, geodesics,
  stability}.
\newblock {\em Grav. Cosmol.} {\bf 1998}, {\em 4},~128--138,
  \href{http://xxx.lanl.gov/abs/gr-qc/9804064}{{\normalfont
  [arXiv:gr-qc/gr-qc/9804064]}}.
\newblock [Submitted to: Grav. Cosmol.(1998)].

\bibitem[Morris and Thorne(1988)]{Morris:1988cz}
Morris, M.S.; Thorne, K.S.
\newblock {Wormholes in space-time and their use for interstellar travel: A
  tool for teaching general relativity}.
\newblock {\em Am. J. Phys.} {\bf 1988}, {\em 56},~395--412.

\bibitem[Heydarzade \em{et~al.}(2015)Heydarzade, Riazi, and
  Moradpour]{Heydarzade:2014ada}
Heydarzade, Y.; Riazi, N.; Moradpour, H.
\newblock {Phantom Wormhole Solutions in a Generic Cosmological Constant
  Background}.
\newblock {\em Can. J. Phys.} {\bf 2015}, {\em 93},~1523--1531,
  \href{http://xxx.lanl.gov/abs/1411.6294}{{\normalfont
  [arXiv:gr-qc/1411.6294]}}.

\bibitem[Lemos \em{et~al.}(2003)Lemos, Lobo, and Quinet~de
  Oliveira]{Lemos:2003jb}
Lemos, J.P.S.; Lobo, F.S.N.; Quinet~de Oliveira, S.
\newblock {Morris-Thorne wormholes with a cosmological constant}.
\newblock {\em Phys. Rev.} {\bf 2003}, {\em D68},~064004,
  \href{http://xxx.lanl.gov/abs/gr-qc/0302049}{{\normalfont
  [arXiv:gr-qc/gr-qc/0302049]}}.

\bibitem[Narahara \em{et~al.}(1994)Narahara, Furihata, and
  Sato]{Narahara:1994np}
Narahara, K.; Furihata, Y.; Sato, K.
\newblock {Traversable wormhole in the expanding universe}.
\newblock {\em Phys. Lett.} {\bf 1994}, {\em B336},~319--323.

\bibitem[Riazi and Esfahani(2000)]{Riazi:2000uu}
Riazi, N.; Esfahani, B.N.
\newblock {Time dependent wormholes in an expanding universe dominated by
  traceless matter}.
\newblock {\em Astrophys. Space Sci.} {\bf 2000}, {\em 271},~237--243.

\bibitem[Hinshaw \em{et~al.}(2013)Hinshaw et~al.]{2013ApJS19H}
Hinshaw, G.; others.
\newblock {Nine-Year Wilkinson Microwave Anisotropy Probe (WMAP) Observations:
  Cosmological Parameter Results}.
\newblock {\em Astrophys.J.Suppl.} {\bf 2013}, {\em 208},~19.

\bibitem[Nojiri and Odintsov(2003)]{Nojiri:2003vn}
Nojiri, S.; Odintsov, S.D.
\newblock {Quantum de Sitter cosmology and phantom matter}.
\newblock {\em Phys. Lett.} {\bf 2003}, {\em B562},~147--152,
  \href{http://xxx.lanl.gov/abs/hep-th/0303117}{{\normalfont
  [arXiv:hep-th/hep-th/0303117]}}.

\bibitem[Carroll \em{et~al.}(2003)Carroll, Hoffman, and Trodden]{Carroll2003st}
Carroll, S.M.; Hoffman, M.; Trodden, M.
\newblock {Can the dark energy equation - of - state parameter w be less than
  -1?}
\newblock {\em Phys.Rev.} {\bf 2003}, {\em D68},~023509,
  \href{http://xxx.lanl.gov/abs/astro-ph/0301273}{{\normalfont
  [arXiv:astro-ph/astro-ph/0301273]}}.

\bibitem[Scheel \em{et~al.}(1995)Scheel, Shapiro, and
  Teukolsky]{PhysRevD.51.4236}
Scheel, M.; Shapiro, S.; Teukolsky, S.
\newblock Collapse to black holes in Brans-Dicke theory. II. Comparison with
  general relativity.
\newblock {\em Phys. Rev. D} {\bf 1995}, {\em 51},~4236--4249.

\bibitem[Uehara and Kim(1982)]{Uehara:1981nq}
Uehara, K.; Kim, C.W.
\newblock {Brans-dicke Cosmology With the Cosmological Constant}.
\newblock {\em Phys. Rev.} {\bf 1982}, {\em D26},~2575.

\bibitem[Riazi and Ahmadi-Azar(1995)]{Riazi:1995ms}
Riazi, N.; Ahmadi-Azar, E.
\newblock {A Class of Exact Cosmological Solutions of Brans-Dicke Theory With
  Cosmological Constant}.
\newblock {\em Astrophys. Space Sci.} {\bf 1995}, {\em 226},~1.

\bibitem[Pimentel(1985)]{Pimentel1985}
Pimentel, L.O.
\newblock Exact cosmological solutions in the scalar-tensor theory with
  cosmological constant.
\newblock {\em Astrophysics and Space Science} {\bf 1985}, {\em 112},~175--183.

\bibitem[Ram and Singh(1997)]{Ram:1997un}
Ram, S.; Singh, C.P.
\newblock {Early cosmological models with bulk viscosity in Brans-Dicke
  theory}.
\newblock {\em Astrophys. Space Sci.} {\bf 1997}, {\em 254},~143--150.

\bibitem[Pandey(2001)]{Pandey:2001ua}
Pandey, S.N.
\newblock {Brans-Dicke cosmology with non-vanishing cosmological constant and
  non-zero curvature}.
\newblock {\em Astrophys. Space Sci.} {\bf 2001}, {\em 277},~403--408.

\bibitem[Romero and Barros(1993)]{Romero:1992ci}
Romero, C.; Barros, A.
\newblock {Brans-Dicke vacuum solutions and the cosmological constant: A
  Qualitative analysis}.
\newblock {\em Gen. Rel. Grav.} {\bf 1993}, {\em 25},~491--502.

\bibitem[Romero and Barros(1992)]{Romero:1992xx}
Romero, C.; Barros, A.
\newblock {Brans-Dicke cosmology and the cosmological constant: the spectrum of
  vacuum solutions}.
\newblock {\em Astrophys. Space Sci.} {\bf 1992}, {\em 192},~263--274.

\bibitem[Cerver{\'o} and Est{\'e}vez(1983)]{Cervero1983}
Cerver{\'o}, J.M.; Est{\'e}vez, P.G.
\newblock General solutions for a cosmological Robertsonwalker metric in the
  Brans-Dicke theory.
\newblock {\em General Relativity and Gravitation} {\bf 1983}, {\em
  15},~351--356.

\bibitem[Bhattacharya \em{et~al.}(2017)Bhattacharya, Dialektopoulos, Romano,
  Skordis, and Tomaras]{Bhattacharya:2016vur}
Bhattacharya, S.; Dialektopoulos, K.F.; Romano, A.E.; Skordis, C.; Tomaras,
  T.N.
\newblock {The maximum sizes of large scale structures in alternative theories
  of gravity}.
\newblock {\em JCAP} {\bf 2017}, {\em 1707},~018,
  \href{http://xxx.lanl.gov/abs/1611.05055}{{\normalfont
  [arXiv:astro-ph.CO/1611.05055]}}.

\bibitem[Xiao and Zhu(1996)]{Xiao:1991nv}
Xiao, X.G.; Zhu, J.Y.
\newblock {Wormhole solution in vacuum Brans-Dicke theory with cosmological
  constant}.
\newblock {\em Chin. Phys. Lett.} {\bf 1996}, {\em 13},~405--408.

\bibitem[Hrycyna and Szydłowski(2013)]{Hrycyna:2013yia}
Hrycyna, O.; Szydłowski, M.
\newblock {Dynamical complexity of the Brans-Dicke cosmology}.
\newblock {\em JCAP} {\bf 2013}, {\em 1312},~016,
  \href{http://xxx.lanl.gov/abs/1310.1961}{{\normalfont
  [arXiv:gr-qc/1310.1961]}}.

\bibitem[Tretyakova \em{et~al.}(2015)Tretyakova, Latosh, and
  Alexeyev]{Tretyakova:2015vaa}
Tretyakova, D.A.; Latosh, B.N.; Alexeyev, S.O.
\newblock {Wormholes and naked singularities in Brans–Dicke cosmology}.
\newblock {\em Class. Quant. Grav.} {\bf 2015}, {\em 32},~185002,
  \href{http://xxx.lanl.gov/abs/1504.06723}{{\normalfont
  [arXiv:gr-qc/1504.06723]}}.

\bibitem[Agnese and La~Camera(2001)]{Agnese:2000dv}
Agnese, A.G.; La~Camera, M.
\newblock {Schwarzschild metrics and quasiuniverses}.
\newblock {\em Found. Phys. Lett.} {\bf 2001}, {\em 14},~581--587,
  \href{http://xxx.lanl.gov/abs/gr-qc/0007041}{{\normalfont
  [arXiv:gr-qc/gr-qc/0007041]}}.

\bibitem[Alexeyev \em{et~al.}(2011)Alexeyev, Rannu, and
  Gareeva]{Alexeyev:2011hc}
Alexeyev, S.O.; Rannu, K.A.; Gareeva, D.V.
\newblock {Possible observation sequences of Brans-Dicke wormholes}.
\newblock {\em J. Exp. Theor. Phys.} {\bf 2011}, {\em 113},~628--636,
  \href{http://xxx.lanl.gov/abs/1104.2536}{{\normalfont
  [arXiv:gr-qc/1104.2536]}}.
\newblock [Zh. Eksp. Teor. Fiz.140,722(2011)].

\bibitem[Horndeski(1974)]{Horndeski:1974wa}
Horndeski, G.W.
\newblock {Second-order scalar-tensor field equations in a four-dimensional
  space}.
\newblock {\em Int. J. Theor. Phys.} {\bf 1974}, {\em 10},~363--384.

\bibitem[Nicolis \em{et~al.}(2009)Nicolis, Rattazzi, and
  Trincherini]{Nicolis:2008in}
Nicolis, A.; Rattazzi, R.; Trincherini, E.
\newblock {The Galileon as a local modification of gravity}.
\newblock {\em Phys. Rev.} {\bf 2009}, {\em D79},~064036,
  \href{http://xxx.lanl.gov/abs/0811.2197}{{\normalfont
  [arXiv:hep-th/0811.2197]}}.

\bibitem[Deffayet \em{et~al.}(2009)Deffayet, Esposito-Far\`ese, and
  Vikman]{Deffayet:2009wt}
Deffayet, C.; Esposito-Far\`ese, G.; Vikman, A.
\newblock {Covariant Galileon}.
\newblock {\em Phys. Rev.} {\bf 2009}, {\em D79},~084003,
  \href{http://xxx.lanl.gov/abs/0901.1314}{{\normalfont
  [arXiv:hep-th/0901.1314]}}.

\bibitem[Kobayashi \em{et~al.}(2011)Kobayashi, Yamaguchi, and
  Yokoyama]{Kobayashi:2011nu}
Kobayashi, T.; Yamaguchi, M.; Yokoyama, J.
\newblock {Generalized G-inflation: Inflation with the most general
  second-order field equations}.
\newblock {\em Prog. Theor. Phys.} {\bf 2011}, {\em 126},~511--529,
  \href{http://xxx.lanl.gov/abs/1105.5723}{{\normalfont
  [arXiv:hep-th/1105.5723]}}.

\bibitem[Deffayet \em{et~al.}(2009)Deffayet, Deser, and
  Esposito-Far\`ese]{Deffayet:2009mn}
Deffayet, C.; Deser, S.; Esposito-Far\`ese, G.
\newblock {Generalized Galileons: All scalar models whose curved background
  extensions maintain second-order field equations and stress-tensors}.
\newblock {\em Phys. Rev.} {\bf 2009}, {\em D80},~064015,
  \href{http://xxx.lanl.gov/abs/0906.1967}{{\normalfont
  [arXiv:gr-qc/0906.1967]}}.

\bibitem[Kobayashi and Tanahashi(2014)]{Kobayashi:2014eva}
Kobayashi, T.; Tanahashi, N.
\newblock {Exact black hole solutions in shift symmetric scalar–tensor
  theories}.
\newblock {\em PTEP} {\bf 2014}, {\em 2014},~073E02,
  \href{http://xxx.lanl.gov/abs/1403.4364}{{\normalfont
  [arXiv:gr-qc/1403.4364]}}.

\bibitem[Starobinsky \em{et~al.}(2016)Starobinsky, Sushkov, and
  Volkov]{Starobinsky:2016kua}
Starobinsky, A.A.; Sushkov, S.V.; Volkov, M.S.
\newblock {The screening Horndeski cosmologies}.
\newblock {\em JCAP} {\bf 2016}, {\em 1606},~007,
  \href{http://xxx.lanl.gov/abs/1604.06085}{{\normalfont
  [arXiv:hep-th/1604.06085]}}.

\bibitem[Charmousis and Iosifidis(2015)]{Charmousis:2015aya}
Charmousis, C.; Iosifidis, D.
\newblock {Self tuning scalar tensor black holes}.
\newblock {\em J. Phys. Conf. Ser.} {\bf 2015}, {\em 600},~012003,
  \href{http://xxx.lanl.gov/abs/1501.05167}{{\normalfont
  [arXiv:gr-qc/1501.05167]}}.

\bibitem[Rinaldi(2012)]{Rinaldi:2012vy}
Rinaldi, M.
\newblock {Black holes with non-minimal derivative coupling}.
\newblock {\em Phys. Rev.} {\bf 2012}, {\em D86},~084048,
  \href{http://xxx.lanl.gov/abs/1208.0103}{{\normalfont
  [arXiv:gr-qc/1208.0103]}}.

\bibitem[Babichev and Charmousis(2014)]{Babichev:2013cya}
Babichev, E.; Charmousis, C.
\newblock {Dressing a black hole with a time-dependent Galileon}.
\newblock {\em JHEP} {\bf 2014}, {\em 08},~106,
  \href{http://xxx.lanl.gov/abs/1312.3204}{{\normalfont
  [arXiv:gr-qc/1312.3204]}}.

\bibitem[Babichev \em{et~al.}(2015)Babichev, Charmousis, and
  Hassaine]{Babichev:2015rva}
Babichev, E.; Charmousis, C.; Hassaine, M.
\newblock {Charged Galileon black holes}.
\newblock {\em JCAP} {\bf 2015}, {\em 1505},~031,
  \href{http://xxx.lanl.gov/abs/1503.02545}{{\normalfont
  [arXiv:gr-qc/1503.02545]}}.

\bibitem[Anabalon \em{et~al.}(2014)Anabalon, Cisterna, and
  Oliva]{Anabalon:2013oea}
Anabalon, A.; Cisterna, A.; Oliva, J.
\newblock {Asymptotically locally AdS and flat black holes in Horndeski
  theory}.
\newblock {\em Phys. Rev.} {\bf 2014}, {\em D89},~084050,
  \href{http://xxx.lanl.gov/abs/1312.3597}{{\normalfont
  [arXiv:gr-qc/1312.3597]}}.

\bibitem[Minamitsuji(2014)]{Minamitsuji:2013ura}
Minamitsuji, M.
\newblock {Solutions in the scalar-tensor theory with nonminimal derivative
  coupling}.
\newblock {\em Phys. Rev.} {\bf 2014}, {\em D89},~064017,
  \href{http://xxx.lanl.gov/abs/1312.3759}{{\normalfont
  [arXiv:gr-qc/1312.3759]}}.

\bibitem[Babichev \em{et~al.}(2016)Babichev, Charmousis, and
  Lehébel]{Babichev:2016rlq}
Babichev, E.; Charmousis, C.; Lehébel, A.
\newblock {Black holes and stars in Horndeski theory}.
\newblock {\em Class. Quant. Grav.} {\bf 2016}, {\em 33},~154002,
  \href{http://xxx.lanl.gov/abs/1604.06402}{{\normalfont
  [arXiv:gr-qc/1604.06402]}}.

\bibitem[Charmousis \em{et~al.}(2014)Charmousis, Kolyvaris, Papantonopoulos,
  and Tsoukalas]{Charmousis:2014zaa}
Charmousis, C.; Kolyvaris, T.; Papantonopoulos, E.; Tsoukalas, M.
\newblock {Black Holes in Bi-scalar Extensions of Horndeski Theories}.
\newblock {\em JHEP} {\bf 2014}, {\em 07},~085,
  \href{http://xxx.lanl.gov/abs/1404.1024}{{\normalfont
  [arXiv:gr-qc/1404.1024]}}.

\bibitem[Tretyakova and Takahashi(2017)]{Tretyakova:2017lyg}
Tretyakova, D.A.; Takahashi, K.
\newblock {Stable black holes in shift-symmetric Horndeski theories}.
\newblock {\em Class. Quant. Grav.} {\bf 2017}, {\em 34},~175007,
  \href{http://xxx.lanl.gov/abs/1702.03502}{{\normalfont
  [arXiv:gr-qc/1702.03502]}}.

\bibitem[Miao and Xu(2016)]{Miao:2016aol}
Miao, Y.G.; Xu, Z.M.
\newblock {Thermodynamics of Horndeski black holes with non-minimal derivative
  coupling}.
\newblock {\em Eur. Phys. J.} {\bf 2016}, {\em C76},~638,
  \href{http://xxx.lanl.gov/abs/1607.06629}{{\normalfont
  [arXiv:hep-th/1607.06629]}}.

\bibitem[Feng \em{et~al.}(2016)Feng, Liu, Lü, and Pope]{Feng:2015wvb}
Feng, X.H.; Liu, H.S.; Lü, H.; Pope, C.N.
\newblock {Thermodynamics of Charged Black Holes in Einstein-Horndeski-Maxwell
  Theory}.
\newblock {\em Phys. Rev.} {\bf 2016}, {\em D93},~044030,
  \href{http://xxx.lanl.gov/abs/1512.02659}{{\normalfont
  [arXiv:hep-th/1512.02659]}}.

\bibitem[Feng \em{et~al.}(2015)Feng, Liu, Lü, and Pope]{Feng:2015oea}
Feng, X.H.; Liu, H.S.; Lü, H.; Pope, C.N.
\newblock {Black Hole Entropy and Viscosity Bound in Horndeski Gravity}.
\newblock {\em JHEP} {\bf 2015}, {\em 11},~176,
  \href{http://xxx.lanl.gov/abs/1509.07142}{{\normalfont
  [arXiv:hep-th/1509.07142]}}.

\bibitem[Sushkov(2017)]{Sushkov:2017ueq}
Sushkov, S.V.
\newblock {Horndeski Wormholes}.
\newblock {\em Fundam. Theor. Phys.} {\bf 2017}, {\em 189},~89--109.

\bibitem[Korolev and Sushkov(2014)]{Korolev:2014hwa}
Korolev, R.V.; Sushkov, S.V.
\newblock {Exact wormhole solutions with nonminimal kinetic coupling}.
\newblock {\em Phys. Rev.} {\bf 2014}, {\em D90},~124025,
  \href{http://xxx.lanl.gov/abs/1408.1235}{{\normalfont
  [arXiv:gr-qc/1408.1235]}}.

\bibitem[Kobayashi \em{et~al.}(2010)Kobayashi, Yamaguchi, and
  Yokoyama]{Kobayashi:2010cm}
Kobayashi, T.; Yamaguchi, M.; Yokoyama, J.
\newblock {G-inflation: Inflation driven by the Galileon field}.
\newblock {\em Phys. Rev. Lett.} {\bf 2010}, {\em 105},~231302,
  \href{http://xxx.lanl.gov/abs/1008.0603}{{\normalfont
  [arXiv:hep-th/1008.0603]}}.

\bibitem[Kanti \em{et~al.}(2015)Kanti, Gannouji, and Dadhich]{Kanti:2015pda}
Kanti, P.; Gannouji, R.; Dadhich, N.
\newblock {Gauss-Bonnet Inflation}.
\newblock {\em Phys. Rev.} {\bf 2015}, {\em D92},~041302,
  \href{http://xxx.lanl.gov/abs/1503.01579}{{\normalfont
  [arXiv:hep-th/1503.01579]}}.

\bibitem[Fomin and Chervon(2017)]{Fomin:2017vae}
Fomin, I.V.; Chervon, S.V.
\newblock {Exact Inflation in Einstein-Gauss-Bonnet Gravity}.
\newblock  {5th Ulyanovsk International School Seminar: Problems of Theoretical
  and Observational Cosmology (UISS 2016) Ulyanovsk, Russia, September 19-30,
  2016},  2017,  \href{http://xxx.lanl.gov/abs/1704.03634}{{\normalfont
  [arXiv:gr-qc/1704.03634]}}.

\bibitem[Babichev \em{et~al.}(2017)Babichev, Charmousis, and
  Lehébel]{Babichev:2017guv}
Babichev, E.; Charmousis, C.; Lehébel, A.
\newblock {Asymptotically flat black holes in Horndeski theory and beyond}.
\newblock {\em JCAP} {\bf 2017}, {\em 1704},~027,
  \href{http://xxx.lanl.gov/abs/1702.01938}{{\normalfont
  [arXiv:gr-qc/1702.01938]}}.

\bibitem[Babichev \em{et~al.}(2016)Babichev, Charmousis, Lehébel, and
  Moskalets]{Babichev:2016fbg}
Babichev, E.; Charmousis, C.; Lehébel, A.; Moskalets, T.
\newblock {Black holes in a cubic Galileon universe}.
\newblock {\em JCAP} {\bf 2016}, {\em 1609},~011,
  \href{http://xxx.lanl.gov/abs/1605.07438}{{\normalfont
  [arXiv:gr-qc/1605.07438]}}.

\bibitem[Moffat(2006)]{Moffat:2005si}
Moffat, J.W.
\newblock {Scalar-tensor-vector gravity theory}.
\newblock {\em JCAP} {\bf 2006}, {\em 0603},~004,
  \href{http://xxx.lanl.gov/abs/gr-qc/0506021}{{\normalfont
  [arXiv:gr-qc/gr-qc/0506021]}}.

\bibitem[Moffat and Rahvar(2013)]{Moffat:2013sja}
Moffat, J.W.; Rahvar, S.
\newblock {The MOG weak field approximation and observational test of galaxy
  rotation curves}.
\newblock {\em Mon. Not. Roy. Astron. Soc.} {\bf 2013}, {\em 436},~1439--1451,
  \href{http://xxx.lanl.gov/abs/1306.6383}{{\normalfont
  [arXiv:astro-ph.GA/1306.6383]}}.

\bibitem[Iorio(2008)]{Iorio:2008sk}
Iorio, L.
\newblock {Putting Yukawa-like Modified Gravity (MOG) on the test in the Solar
  System}.
\newblock {\em Schol. Res. Exch.} {\bf 2008}, {\em 2008},~238385,
  \href{http://xxx.lanl.gov/abs/0809.3563}{{\normalfont
  [arXiv:gr-qc/0809.3563]}}.

\bibitem[Iorio(2002)]{Iorio:2002jy}
Iorio, L.
\newblock {Constraints to a Yukawa gravitational potential from laser data to
  LAGEOS satellites}.
\newblock {\em Phys. Lett.} {\bf 2002}, {\em A298},~315--318,
  \href{http://xxx.lanl.gov/abs/gr-qc/0201081}{{\normalfont
  [arXiv:gr-qc/gr-qc/0201081]}}.

\bibitem[Iorio(2007)]{Iorio:2007gq}
Iorio, L.
\newblock {Constraints on the range lambda of Yukawa-like modifications to the
  Newtonian inverse-square law of gravitation from Solar System planetary
  motions}.
\newblock {\em JHEP} {\bf 2007}, {\em 10},~041,
  \href{http://xxx.lanl.gov/abs/0708.1080}{{\normalfont
  [arXiv:gr-qc/0708.1080]}}.

\bibitem[Moffat and Toth(2013)]{Moffat:2011rp}
Moffat, J.W.; Toth, V.T.
\newblock {Cosmological observations in a modified theory of gravity (MOG)}.
\newblock {\em Galaxies} {\bf 2013}, {\em 1},~65--82,
  \href{http://xxx.lanl.gov/abs/1104.2957}{{\normalfont
  [arXiv:astro-ph.CO/1104.2957]}}.

\bibitem[Moffat \em{et~al.}(2012)Moffat, Rahvar, and Toth]{Moffat:2012wn}
Moffat, J.W.; Rahvar, S.; Toth, V.T.
\newblock {Applying MOG to lensing: Einstein rings, Abell 520 and the Bullet
  Cluster} {\bf 2012}.
\newblock  \href{http://xxx.lanl.gov/abs/1204.2985}{{\normalfont
  [arXiv:astro-ph.CO/1204.2985]}}.

\bibitem[Moffat and Toth(2007)]{Moffat:2007qv}
Moffat, J.W.; Toth, V.T.
\newblock {Testing modified gravity with motion of satellites around galaxies}
  {\bf 2007}.
\newblock  \href{http://xxx.lanl.gov/abs/0708.1264}{{\normalfont
  [arXiv:astro-ph/0708.1264]}}.

\bibitem[Moffat and Toth(2009)]{Moffat:2008gi}
Moffat, J.W.; Toth, V.T.
\newblock {The Bending of light and lensing in modified gravity}.
\newblock {\em Mon. Not. Roy. Astron. Soc.} {\bf 2009}, {\em 397},~1885--1992,
  \href{http://xxx.lanl.gov/abs/0805.4774}{{\normalfont
  [arXiv:astro-ph/0805.4774]}}.

\bibitem[Moffat(2006)]{Moffat:2006gz}
Moffat, J.W.
\newblock {Time delay predictions in a modified gravity theory}.
\newblock {\em Class. Quant. Grav.} {\bf 2006}, {\em 23},~6767--6772,
  \href{http://xxx.lanl.gov/abs/gr-qc/0605141}{{\normalfont
  [arXiv:gr-qc/gr-qc/0605141]}}.

\bibitem[Moffat and Toth(2009)]{Moffat:2007nj}
Moffat, J.W.; Toth, V.T.
\newblock {Fundamental parameter-free solutions in modified gravity}.
\newblock {\em Class. Quant. Grav.} {\bf 2009}, {\em 26},~085002,
  \href{http://xxx.lanl.gov/abs/0712.1796}{{\normalfont
  [arXiv:gr-qc/0712.1796]}}.

\bibitem[Moffat(2008)]{Moffat:2006rq}
Moffat, J.W.
\newblock {A Modified Gravity and its Consequences for the Solar System,
  Astrophysics and Cosmology}.
\newblock {\em Int. J. Mod. Phys.} {\bf 2008}, {\em D16},~2075--2090,
  \href{http://xxx.lanl.gov/abs/gr-qc/0608074}{{\normalfont
  [arXiv:gr-qc/gr-qc/0608074]}}.

\bibitem[Moffat(2015)]{Moffat:2014aja}
Moffat, J.W.
\newblock {Black Holes in Modified Gravity (MOG)}.
\newblock {\em Eur. Phys. J.} {\bf 2015}, {\em C75},~175,
  \href{http://xxx.lanl.gov/abs/1412.5424}{{\normalfont
  [arXiv:gr-qc/1412.5424]}}.

\bibitem[Mureika \em{et~al.}(2016)Mureika, Moffat, and Faizal]{Mureika:2015sda}
Mureika, J.R.; Moffat, J.W.; Faizal, M.
\newblock {Black hole thermodynamics in MOdified Gravity (MOG)}.
\newblock {\em Phys. Lett.} {\bf 2016}, {\em B757},~528--536,
  \href{http://xxx.lanl.gov/abs/1504.08226}{{\normalfont
  [arXiv:gr-qc/1504.08226]}}.

\bibitem[Ayon-Beato and Garcia(1998)]{AyonBeato:1998ub}
Ayon-Beato, E.; Garcia, A.
\newblock {Regular black hole in general relativity coupled to nonlinear
  electrodynamics}.
\newblock {\em Phys. Rev. Lett.} {\bf 1998}, {\em 80},~5056--5059,
  \href{http://xxx.lanl.gov/abs/gr-qc/9911046}{{\normalfont
  [arXiv:gr-qc/gr-qc/9911046]}}.

\bibitem[Ayon-Beato and Garcia(1999)]{AyonBeato:1999ec}
Ayon-Beato, E.; Garcia, A.
\newblock {Nonsingular charged black hole solution for nonlinear source}.
\newblock {\em Gen. Rel. Grav.} {\bf 1999}, {\em 31},~629--633,
  \href{http://xxx.lanl.gov/abs/gr-qc/9911084}{{\normalfont
  [arXiv:gr-qc/gr-qc/9911084]}}.

\bibitem[Skordis(2009)]{Skordis:2009bf}
Skordis, C.
\newblock {The Tensor-Vector-Scalar theory and its cosmology}.
\newblock {\em Class. Quant. Grav.} {\bf 2009}, {\em 26},~143001,
  \href{http://xxx.lanl.gov/abs/0903.3602}{{\normalfont
  [arXiv:astro-ph.CO/0903.3602]}}.

\bibitem[Bekenstein(2004)]{Bekenstein:2004ne}
Bekenstein, J.D.
\newblock {Relativistic gravitation theory for the MOND paradigm}.
\newblock {\em Phys. Rev.} {\bf 2004}, {\em D70},~083509,
  \href{http://xxx.lanl.gov/abs/astro-ph/0403694}{{\normalfont
  [arXiv:astro-ph/astro-ph/0403694]}}.
\newblock [Erratum: Phys. Rev.D71,069901(2005)].

\bibitem[Zhao(2008)]{Zhao:2006vm}
Zhao, H.
\newblock {Constraining TeVeS Gravity as Effective Dark Matter and Dark
  Energy}.
\newblock {\em Int. J. Mod. Phys.} {\bf 2008}, {\em D16},~2055--2063,
  \href{http://xxx.lanl.gov/abs/astro-ph/0610056}{{\normalfont
  [arXiv:astro-ph/astro-ph/0610056]}}.

\bibitem[Armendariz-Picon \em{et~al.}(1999)Armendariz-Picon, Damour, and
  Mukhanov]{ArmendarizPicon:1999rj}
Armendariz-Picon, C.; Damour, T.; Mukhanov, V.F.
\newblock {k - inflation}.
\newblock {\em Phys. Lett.} {\bf 1999}, {\em B458},~209--218,
  \href{http://xxx.lanl.gov/abs/hep-th/9904075}{{\normalfont
  [arXiv:hep-th/hep-th/9904075]}}.

\bibitem[Armendariz-Picon \em{et~al.}(2000)Armendariz-Picon, Mukhanov, and
  Steinhardt]{ArmendarizPicon:2000dh}
Armendariz-Picon, C.; Mukhanov, V.F.; Steinhardt, P.J.
\newblock {A Dynamical solution to the problem of a small cosmological constant
  and late time cosmic acceleration}.
\newblock {\em Phys. Rev. Lett.} {\bf 2000}, {\em 85},~4438--4441,
  \href{http://xxx.lanl.gov/abs/astro-ph/0004134}{{\normalfont
  [arXiv:astro-ph/astro-ph/0004134]}}.

\bibitem[Giannios(2005)]{Giannios:2005es}
Giannios, D.
\newblock {Spherically symmetric, static spacetimes in TeVeS}.
\newblock {\em Phys. Rev.} {\bf 2005}, {\em D71},~103511,
  \href{http://xxx.lanl.gov/abs/gr-qc/0502122}{{\normalfont
  [arXiv:gr-qc/gr-qc/0502122]}}.

\bibitem[Sagi and Bekenstein(2008)]{Sagi:2007hb}
Sagi, E.; Bekenstein, J.D.
\newblock {Black holes in the TeVeS theory of gravity and their
  thermodynamics}.
\newblock {\em Phys. Rev.} {\bf 2008}, {\em D77},~024010,
  \href{http://xxx.lanl.gov/abs/0708.2639}{{\normalfont
  [arXiv:gr-qc/0708.2639]}}.

\bibitem[Skordis and Zlosnik(2012)]{Skordis:2011ad}
Skordis, C.; Zlosnik, T.
\newblock {The Geometry Of Modified Newtonian Dynamics}.
\newblock {\em Phys. Rev.} {\bf 2012}, {\em D85},~044044,
  \href{http://xxx.lanl.gov/abs/1101.6019}{{\normalfont
  [arXiv:gr-qc/1101.6019]}}.

\bibitem[Blas and Lim(2015)]{Blas:2014aca}
Blas, D.; Lim, E.
\newblock {Phenomenology of theories of gravity without Lorentz invariance: the
  preferred frame case}.
\newblock {\em Int. J. Mod. Phys.} {\bf 2015}, {\em D23},~1443009,
  \href{http://xxx.lanl.gov/abs/1412.4828}{{\normalfont
  [arXiv:gr-qc/1412.4828]}}.

\bibitem[Skenderis(2000)]{Skenderis:1999bs}
Skenderis, K.
\newblock {Black holes and branes in string theory}.
\newblock {\em Lect. Notes Phys.} {\bf 2000}, {\em 541},~325--364,
  \href{http://xxx.lanl.gov/abs/hep-th/9901050}{{\normalfont
  [arXiv:hep-th/hep-th/9901050]}}.
\newblock [,345(1999)].

\bibitem[Mathur(2005)]{Mathur:2005zp}
Mathur, S.D.
\newblock {The Fuzzball proposal for black holes: An Elementary review}.
\newblock {\em Fortsch. Phys.} {\bf 2005}, {\em 53},~793--827,
  \href{http://xxx.lanl.gov/abs/hep-th/0502050}{{\normalfont
  [arXiv:hep-th/hep-th/0502050]}}.

\bibitem[Susskind(1993)]{Susskind:1993ws}
Susskind, L.
\newblock {Some speculations about black hole entropy in string theory} {\bf
  1993}.
\newblock  \href{http://xxx.lanl.gov/abs/hep-th/9309145}{{\normalfont
  [arXiv:hep-th/hep-th/9309145]}}.

\bibitem[Russo and Susskind(1995)]{Russo:1994ev}
Russo, J.G.; Susskind, L.
\newblock {Asymptotic level density in heterotic string theory and rotating
  black holes}.
\newblock {\em Nucl. Phys.} {\bf 1995}, {\em B437},~611--626,
  \href{http://xxx.lanl.gov/abs/hep-th/9405117}{{\normalfont
  [arXiv:hep-th/hep-th/9405117]}}.

\bibitem[Skenderis and Taylor(2008)]{Skenderis:2008qn}
Skenderis, K.; Taylor, M.
\newblock {The fuzzball proposal for black holes}.
\newblock {\em Phys. Rept.} {\bf 2008}, {\em 467},~117--171,
  \href{http://xxx.lanl.gov/abs/0804.0552}{{\normalfont
  [arXiv:hep-th/0804.0552]}}.

\bibitem[Donoghue(1994)]{Donoghue:1994dn}
Donoghue, J.F.
\newblock {General relativity as an effective field theory: The leading quantum
  corrections}.
\newblock {\em Phys. Rev.} {\bf 1994}, {\em D50},~3874--3888,
  \href{http://xxx.lanl.gov/abs/gr-qc/9405057}{{\normalfont
  [arXiv:gr-qc/gr-qc/9405057]}}.

\bibitem[Burgess(2004)]{Burgess:2003jk}
Burgess, C.P.
\newblock {Quantum gravity in everyday life: General relativity as an effective
  field theory}.
\newblock {\em Living Rev. Rel.} {\bf 2004}, {\em 7},~5--56,
  \href{http://xxx.lanl.gov/abs/gr-qc/0311082}{{\normalfont
  [arXiv:gr-qc/gr-qc/0311082]}}.

\bibitem[Horowitz \em{et~al.}(1996)Horowitz, Maldacena, and
  Strominger]{Horowitz:1996ay}
Horowitz, G.T.; Maldacena, J.M.; Strominger, A.
\newblock {Nonextremal black hole microstates and U duality}.
\newblock {\em Phys. Lett.} {\bf 1996}, {\em B383},~151--159,
  \href{http://xxx.lanl.gov/abs/hep-th/9603109}{{\normalfont
  [arXiv:hep-th/hep-th/9603109]}}.

\bibitem[Youm(1999)]{Youm:1997hw}
Youm, D.
\newblock {Black holes and solitons in string theory}.
\newblock {\em Phys. Rept.} {\bf 1999}, {\em 316},~1--232,
  \href{http://xxx.lanl.gov/abs/hep-th/9710046}{{\normalfont
  [arXiv:hep-th/hep-th/9710046]}}.

\bibitem[Cvetic(1997)]{Cvetic:1996uf}
Cvetic, M.
\newblock {Properties of black holes in toroidally compactified string theory}.
\newblock {\em Nucl. Phys. Proc. Suppl.} {\bf 1997}, {\em 56B},~1--10,
  \href{http://xxx.lanl.gov/abs/hep-th/9701152}{{\normalfont
  [arXiv:hep-th/hep-th/9701152]}}.

\end{thebibliography}

\end{document}